\documentclass[utf8]{FrontiersinHarvard} % for articles in journals using the Harvard Referencing Style (Author-Date), for Frontiers Reference Styles by Journal: https://zendesk.frontiersin.org/hc/en-us/articles/360017860337-Frontiers-Reference-Styles-by-Journal
\usepackage{url,hyperref,lineno,microtype,subcaption}
\usepackage[onehalfspacing]{setspace}
\usepackage{aas_macros}

\def\keyFont{\fontsize{8}{11}\helveticabold }
\def\firstAuthorLast{Tricco} %use et al only if is more than 1 author
\def\Authors{Terrence S. Tricco\,$^{*}$}
\def\Address{Department of Computer Science, Memorial University of Newfoundland, St. John's, NL, Canada}
\def\corrAuthor{Terrence S. Tricco}

\def\corrEmail{tstricco@mun.ca}

\usepackage{bm}

\begin{document}
\onecolumn
\firstpage{1}

\title {Smoothed particle magnetohydrodynamics} 

\author[\firstAuthorLast ]{\Authors} %This field will be automatically populated
\address{} %This field will be automatically populated
\correspondance{} %This field will be automatically populated

\extraAuth{}% If there are more than 1 corresponding author, comment this line and uncomment the next one.
%\extraAuth{corresponding Author2 \\ Laboratory X2, Institute X2, Department X2, Organization X2, Street X2, City X2 , State XX2 (only USA, Canada and Australia), Zip Code2, X2 Country X2, email2@uni2.edu}

\twocolumn
\maketitle

\begin{abstract}

\section{}
Smoothed particle magnetohydrodynamics has reached a level of maturity that enables the study of a wide range of astrophysical problems. In this review, the numerical details of the modern SPMHD method are described. The three fundamental components of SPMHD are methods to evolve the magnetic field in time, calculate accelerations from the magnetic field, and maintain the divergence-free constraint on the magnetic field (no monopoles). The connection between these three requirements in SPMHD will be highlighted throughout. The focus of this review is on the methods that work well in practice, with discussion on why they work well and other approaches do not. Numerical instabilities will be discussed, as well as strategies to overcome them. The inclusion of non-ideal MHD effects will be presented. A prospective outlook on possible avenues for further improvements will be discussed.

\tiny
 \keyFont{ \section{Keywords:} smoothed particle magnetohydrodynamics, smoothed particle hydrodynamics, magnetohydrodynamics (MHD), Lagrangian particle methods, non-ideal MHD, divergence cleaning, astrophysics} 
\end{abstract}

\section{Introduction}

Smoothed particle magnetohydrodynamics (SPMHD) is a robust numerical method for solving the equations of magnetohydrodynamics (MHD). It is a Lagrangian, mesh-free method that builds upon the smoothed particle hydrodynamics (SPH) framework \citep{lucy77, gm77}. The general picture of SPH is to solve the equations of hydrodynamics by discretising a fluid into a collection of particles that mimic fluid behaviour. Recent reviews of the fundamentals of SPH include \citet{rosswog09}, \citet{springel10} and \citet{price12}.

SPH, and by extension SPMHD, has many advantages for astrophysics. One, the resolution is tied to the mass. Regions of higher mass have more particles, thus more resolution, which is advantageous as the densest areas are typically the most interesting (e.g., stars forming in a molecular cloud). Two, it is trivial to incorporate gravitational N-body methods since SPH is a particle-based scheme. Three, advection is done perfectly, that is, without any dissipation, since it is a Lagrangian method. Four, it can easily handle complex geometries. Five, the Courant timestep does not depend upon the fluid velocity, thus allowing larger timesteps. And six, perhaps its strongest attribute, it has exact simultaneous conservation of mass, momentum, angular momentum, energy, and entropy to the precision of the time-stepping algorithm. This makes SPH significantly robust and stable since it reflects the conservation properties of nature.

Over the past decade, SPMHD has been used to simulate the evolution and impact of magnetic fields in a wide variety of astrophysical problems, such as the study of single and binary star formation \citep{burzle-etal11a, burzle-etal11b, ptb12, btp14, lbp15, tsukamoto-etal15a, tsukamoto-etal15b, wpb16, wpb17, lb17, tsukamoto-etal18, wbp18a, wbp18b, wbp18c, tsukamoto-etal20, wbb21, wurster-etal22}, star cluster formation \citep{wbp19, dw21}, star formation rates in spiral galaxies \citep{hdb23}, accretion discs \citep{fpb17}, tidal disruption events \citep{bonnerot-etal17}, the magnetic field structure of spiral galaxies \citep{dobbs-etal16, ws23}, and galaxy cluster formation \citep{bkw12, barnes-etal18}. It has been shown to yield correct behaviour for the small-scale dynamo amplification of magnetic fields \citep{tpf16}, and can sustain turbulence incited by the magnetorotational instability (MRI) \citep{deng-etal19, wissing-etal22}. SPMHD has also found use outside of astrophysics, where it has been employed to study pinch plasmas and fusion \citep{vela-vela-etal19, tc20, park-etal23}.

The minimum requirements for incorporating MHD into SPH are three-fold: (1) an approach for evolving the magnetic field forward in time, (2) calculation of the accelerations deriving from the Lorentz force, and (3) an approach to uphold the divergence-free constraint of the magnetic field. These need not be independent. Adhering to the divergence-free constraint is strongly connected to the choice of how the magnetic field is numerically evolved, for instance.

The modern SPMHD method solves the set of continuum equations given by
\begin{align}
\frac{{\rm d}\rho}{{\rm d}t} &= - \rho \frac{\partial v^i}{\partial x^i} \label{eq:continuity} , \\
\frac{{\rm d}v^i}{{\rm d}t} &= \frac{1}{\rho} \frac{\partial S^{\ij}}{\partial x^j} \label{eq:mhdmomentumeqn} , \\
\frac{{\rm d}}{{\rm d}t} \left( \frac{B^i}{\rho} \right)  &= \frac{B^j}{\rho}  \frac{\partial v^i}{\partial x^j} \label{eq:induction} , \\
\frac{\partial B^i}{\partial x^i} &= 0 , \label{eq:divbconstraint}
\end{align}
where the stress $S^{ij}$ is defined as
\begin{equation}
S^{ij} = -P \delta^{ij} + \frac{1}{\mu_0} \left( B^i B^j - \frac{B^2}{2} \delta^{ij} \right) , \label{eq:stresstensor}
\end{equation}
and $\rho$ is the density, $v$ is the velocity, $B$ is the magnetic field, $P$ is the pressure, and $\mu_0$ is the permeability of free space. This system of equations is closed by a suitable equation of state. Note that the continuum equations are written with Lagrangian derivatives, ${\rm d}/{\rm d}t \equiv \partial / \partial t + v^j \partial / \partial x^j$, since SPMHD is a Lagrangian method. 

The first attempts to include magnetic fields in SPH were performed by \citet{gm77} who considered magnetic polytropes, though in a form which did not conserve momentum or angular momentum. The modern SPMHD method has its roots in the work by \citet{pm85}, who formulated equations of motion that conserve momentum by using the stress tensor, and applied the method to three-dimensional simulations of gravitationally collapsing gas clouds \citep{phillips86a, phillips86b}. 

The challenge with the conservative approach is that it is numerically unstable when the magnetic pressure exceeds the thermodynamic pressure, that is, for plasma $\beta < 1$ where $\beta \equiv P / (B^2 / 2 \mu_0)$. Though the onset of instability is described by this criterion, the fundamental nature of the instability is linked to magnetic accelerations which are not perpendicular to the magnetic field, that is, that originate from magnetic monopoles. Solutions to this instability invariably incur some penalty to the conservation of momentum \citep{mwd95, morris96thesis, bot01}, though at a cost related to the magnitude of divergence errors in the magnetic field. Again, the linkage between calculating the accelerations deriving from the magnetic field and the approach used to uphold the divergence-free constraint of the magnetic field should be noted. 

Carefully evolving the magnetic field in such a way as to avoid divergence errors altogether would be ideal. For grid codes, this can be accomplished with, for example, constrained transport \citep{eh88}, but a significant impediment to do this in SPMHD is the lack of any well-defined surfaces. To date, no satisfactory approach to ensure a truly divergence-free magnetic field has yet to be found for SPMHD \citep{price10, brandenburg10}. Instead, modern SPMHD straightforwardly discretises the induction equation, with the recognition that divergence errors may be generated as there is no intrinsic divergence control. The magnetic field must be corrected through some other procedure. In practice, ``divergence cleaning'' \citep{tp12, tpb16} works well to remove divergence errors and is the method used in many modern SPMHD calculations.

Other pieces of the SPMHD puzzle that had to be solved include construction of the fully conservative equations incorporating varying resolution and magnetic discontinuity capturing terms \citep{pm04a, pm04b, pm05}. It worthy to note that modern SPMHD is equivalent in formulation to the eight-wave approach of \citet{powell94, powell-etal99}, in that the SPMHD equations have source terms related to the divergence of the magnetic field, such that divergence errors are advected with the fluid flow. Additionally, non-ideal MHD formulations for Ohmic dissipation, ambipolar diffusion, and the Hall effect have been constructed \citep{wpa14, wpb16}. SPMHD has also been adapted for 2D axisymmetric geometry, promising computational efficiency for phenomena involving magnetic fields with axial geometries \citep{garcia-senz-etal23}.

The outline of this review is as follows. A brief reminder of the fundamental SPH equations is provided first (section~\ref{sec:sph}). The ingredients comprising modern SPMHD are subsequently discussed, focusing on the three main requirements of approaches to evolve the magnetic field (section~\ref{sec:induction}), to calculate accelerations from the magnetic field (section~\ref{sec:accelerations}), and to maintain the divergence-free constraint of the magnetic field (section~\ref{sec:divb}). Non-ideal extensions are described in section~\ref{sec:non-ideal}, and a prospective outlook is given in section~\ref{sec:future}.

\section{Smoothed particle hydrodynamics}
\label{sec:sph}

In this review, a standard form of SPH is assumed whereby the density of an SPH particle is calculated by mass-weighted summation. A number of SPH variants and implementations have been proposed over time \citep{rt01, hopkins13, sm13, gsce22}, but here we focus on traditional mass-weighted SPH. The mass-weighted density summation is given by
\begin{equation}
\rho_a = \sum_b m_b W_{ab}(h_a) , \label{eq:dens-sum}
\end{equation}
with $h$ the smoothing length, $m$ the mass, and summation is over neighbouring particles within the support radius of the smoothing kernel, $W_{ab}(h_a) \equiv W(\vert {\bf r}_a - {\bf r}_b \vert, h_a)$, where ${\bf r}$ is the particle coordinates. One could discretise Equation~(\ref{eq:continuity}) directly, though a benefit of Equation~(\ref{eq:dens-sum}) is that the density can then be computed at any location at any time. There is no need to time integrate the density. This also avoids assumptions about the differentiability of the density. See also \citet{price12} for a nuanced presentation on the trickle down effects of using this SPH density estimate. 

Smoothing lengths that are individual per particle can be obtained by simultaneously solving the density summation (Equation~\ref{eq:dens-sum}) with an expression for density given by
\begin{equation}
\rho_a = m_a \left( \frac{h_a}{n_{\rm h}} \right)^{- {\rm n_{dim}}} , \label{eq:dens-estimate}
\end{equation}
where ${\rm n_{dim}}$ is the number of dimensions and $n_{\rm h}$ is a dimensionless quantity specifying the ratio of smoothing length to particle spacing. The smoothing length that yields agreement between Equations~(\ref{eq:dens-sum}) and (\ref{eq:dens-estimate}) can be obtained via a root-finding procedure.

The conservative equations of motion derived from the Lagrangian for the discretised system need only specify the density summation in Equation~(\ref{eq:dens-sum}) \citep{sh02, monaghan02}. Doing so yields
\begin{align}
\frac{{\rm d}\textbf{v}_a}{{\rm d}t} = & - \sum_b m_b \bigg[ \frac{P_a}{\Omega_a \rho_a^2} \nabla_a W_{ab}(h_a) \nonumber \\
& + \frac{P_b}{\Omega_b \rho_b^2} \nabla_a W_{ab}(h_b) \bigg] , \label{eq:sph-momentum}
\end{align}
where factors
\begin{equation}
\Omega_a = 1 - \sum_b m_b \frac{\partial W_{ab}(h_a)}{\partial h_a} \frac{\partial h_a}{\partial \rho_a} 
\end{equation}
are present to account for spatially varying smoothing lengths. Obtaining $\partial h_a / \partial \rho_a$ may be done through Equation~(\ref{eq:dens-estimate}). See \citet{monaghan05, springel10, price12} for detailed derivations of the SPH equations of motion from the Lagrangian.

If the pressure is a function of internal energy, $u$, then a suitable equation must be used to evolve the internal energy forward in time. This can be derived from the first law of thermodynamics, yielding
\begin{equation}
\frac{{\rm d}u}{{\rm d}t} = \frac{P}{\rho^2} \frac{{\rm d}\rho}{{\rm d}t} .
\end{equation}
The time derivative of $\rho$ can be obtained by taking the time derivative of the density summation, Equation~(\ref{eq:dens-sum}). Using this, the discretised internal energy equation is
\begin{equation}
\frac{{\rm d}u_a}{{\rm d}t} = \frac{P_a}{\Omega_a \rho_a^2} \sum_b m_b \textbf{v}_{ab} \cdot \nabla_a W_{ab}(h_a) .
\end{equation}
Note that the total specific energy or entropy of each particle could be evolved instead. The differences between these choices are minimal \citep[see][]{price12}.

The choice of smoothing kernel has many considerations. In practice, the two families of kernels that are most widely used for astrophysical SPMHD calculations are the bell-shaped B-splines and the Wendland kernels. The kernel function can be written in functional form according to
\begin{equation}
W_{ab}(h_a) = \frac{\sigma}{h^{\rm n_{dim}}} w(q) ,
\end{equation}
where $\sigma$ is the normalisation and $q = r_{ab}/h_a$. 
The cubic spline is given by
\begin{equation}
w(q) =
\begin{cases}
(2 - q)^3 - 4 (1 - q)^3 & q < 1, \\
(2 - q)^3  & 1 \leq q < 2 , \\
0 & q \geq 2,
\end{cases}
\end{equation}
with $\sigma = 1/\pi$ in 3D, the quintic spline by
\begin{equation}
w(q) = 
\begin{cases}
(3 - q)^5 - 6 (2 - q)^5 + 15 (1 - q)^5 & q < 1, \\
(3 - q)^5 - 6 (2 - q)^5 & 1 \leq q < 2 , \\
(3 - q)^5  & 2 \leq q < 3 , \\
0 & q \geq 3 ,
\end{cases}
\end{equation}
with $\sigma = 1 / (120 \pi)$ in 3D and Wendland C4 kernel, scaled to a radius of $2h$, defined as
\begin{equation}
w(q) = 
\begin{cases}
\left(1 - q / 2 \right)^6 \left( 35 q^2 / 12 + 3q + 1 \right) & q < 2, \\
0 & q \geq 2 ,
\end{cases}
\end{equation}
with $\sigma = 495 / (256 \pi)$.
The cubic spline has a long history within SPH since it is the lowest order spline that has a continuous first derivative. The quintic spline (or other higher order splines) may be used when higher accuracy is required and the kernel bias needs to be reduced. The B-splines cannot be scaled to arbitrarily large radii because doing so will lead to the formation of close particle pairs. The Wendland kernels are attractive because they are stable against particle pairing for all neighbour numbers owing to their positive definite Fourier transform \citep{da12}. The price they pay for this is that they yield larger density errors than the B-splines at lower neighbour numbers. The choice of kernel plays a role in numerical convergence and the reduction of `E0' errors from the pressure gradient \citep{rha10, mlp12, da12, hopkins15}, but this is dependent upon the particular details of a calculation. In the context of SPMHD, both the cubic spline and Wendland C4 kernel have been used successfully within a variety of dynamo and astrophysical applications, so the choice of kernel seems less important than other numerical details.

\section{Evolving the magnetic field}
\label{sec:induction}

The approach that works best for evolving the magnetic field forward in time is to directly discretise the induction equation. This is also the most straightforward. Once the time derivative of the magnetic field has been calculated, it can be used within standard time integration schemes. A popular option is a kick-drift-kick leapfrog scheme owing to its efficiency, simplicity and conservation properties \citep{gadget2, gasoline2, phantom}.

\subsection{Induction equation}

Discretising the induction equation is as simple as applying standard differencing operators to Equation~(\ref{eq:induction}). Each particle evolves $\textbf{B}/\rho$ according to
\begin{equation}
\label{eq:spmhd-ind-brho}
\frac{{\rm d}}{{\rm d}t} \left( \frac{\textbf{B}}{\rho} \right)_a =  - \frac{1}{\Omega_a \rho_a^2} \sum_b m_b \textbf{v}_{ab} \left[ \textbf{B}_a \cdot \nabla_a W_{ab}(h_a) \right] , 
\end{equation}
where $\textbf{v}_{ab} = \textbf{v}_a - \textbf{v}_b$. The magnetic field, when needed, can simply be reconstructed by multiplying the evolved quantity by the density. 

One could evolve $\textbf{B}$ directly instead of $\textbf{B}/\rho$. Expanding the left-hand side of Equation~(\ref{eq:induction}) yields
\begin{equation}
\frac{{\rm d}\textbf{B}}{{\rm d}t} = \rho \frac{{\rm d}}{{\rm d}t} \left( \frac{\textbf{B}}{\rho} \right) + \frac{\textbf{B}}{\rho} \frac{{\rm d}\rho}{{\rm d}t} , \label{eq:derive-ind-b}
\end{equation}
and by making use of Equation~(\ref{eq:induction}) and the continuity equation (Equation~\ref{eq:continuity}), 
\begin{equation}
\frac{{\rm d}\textbf{B}}{{\rm d}t} = (\textbf{B} \cdot \nabla) \textbf{v} - (\nabla \cdot \textbf{v}) \textbf{B} . \label{eq:ind-dbdt5}
\end{equation}
The corresponding discretisation is
\begin{align}
\frac{{\rm d}\textbf{B}_a}{{\rm d}t} = & - \frac{1}{\Omega_a \rho_a} \sum_b m_b \textbf{v}_{ab} \left[ \textbf{B}_a \cdot \nabla_a W_{ab}(h_a) \right] \nonumber \\
& + \frac{1}{\rho_a} \sum_b m_b \textbf{B}_a \left[ \textbf{v}_{ab} \cdot \nabla_a W_{ab}(h_a) \right] . \label{eq:spmhd-ind-b}
\end{align}
Numerically, any difference between Equations~(\ref{eq:spmhd-ind-brho}) and (\ref{eq:spmhd-ind-b}) should arise solely by the addition of the ${\rm d}\rho/{\rm d}t$ term when evolving $\textbf{B}$ directly. The differential form of the continuity equation is baked into the evolution of $\textbf{B}$, whereas evolving $\textbf{B}/\rho$ uses the integral formulation of the continuity equation (i.e., the density summation) to reconstruct $\textbf{B}$. In theory, evolving $\textbf{B}/\rho$ should thus be preferable over $\textbf{B}$ in situations where a discontinuity is present in the density as it avoids assumptions about the differentiability of the density \citep[see also][]{price08}. In practice, however, no substantive evidence has been found of any meaningful difference between evolving $\textbf{B}/\rho$ and $\textbf{B}$ \citep[e.g.,][]{morris96thesis}

An advantageous property of the above discreti\-sations, aside from their simplicity, is that they are Galilean invariant. That is, the addition of any constant background velocity does not introduce numerical error.

The most significant disadvantage of these discretisations is that they do not place any guarantee on the divergence of the magnetic field. Even with an initially divergence-free magnetic field, numerical errors will lead to divergence errors that can grow over time. A secondary approach is required to treat these errors. The most effective approach to accomplish this at present seems to be mixed hyperbolic/parabolic divergence cleaning \citep{tp12, tpb16}, as is discussed further in section~\ref{sec:divb}.

Additionally, it can be seen that the induction equation solved in SPMHD mimics the source term approach of of \citet{powell94, powell-etal99}. Consider the conservative formulation of the induction equation given by 
\begin{equation}
\frac{\partial \textbf{B}}{\partial t} = \nabla \times (\textbf{v} \times \textbf{B}) . \label{eq:ind-partial}
\end{equation}
The vector calculus identity $\nabla \times (\textbf{v} \times \textbf{B}) \equiv  (\textbf{B} \cdot \nabla) \textbf{v} - (\nabla \cdot \textbf{v}) \textbf{B} - (\textbf{v} \cdot \nabla) \textbf{B} + (\nabla \cdot \textbf{B}) \textbf{v}$ can be used to rewrite Equation~(\ref{eq:ind-partial}) in terms of a Lagrangian derivative, yielding
\begin{equation}
\frac{{\rm d}\textbf{B}}{{\rm d}t} = (\textbf{B} \cdot \nabla) \textbf{v} - (\nabla \cdot \textbf{v}) \textbf{B}  + (\nabla \cdot \textbf{B}) \textbf{v} .
\label{eq:ind-dbdt6}
\end{equation}
This is identical to Equation~(\ref{eq:ind-dbdt5}) except for the term proportional to $\nabla \cdot \textbf{B}$. In SPMHD, this term is ignored, in essence evolving the magnetic field under the assumption that the divergence of the magnetic field is zero. This treatment of monopoles is equivalent to how monopoles are treated by the eight-wave formulation. 

The inclusion or absence of the $\nabla \cdot \textbf{B}$ term consequently means that divergence errors are either, respectively, advected or dispersed. Taking the divergence of Equation~(\ref{eq:ind-dbdt6}) yields,
\begin{equation}
\frac{\partial (\nabla \cdot \textbf{B})}{\partial t} = 0,
\label{eq:ind-eulerian}
\end{equation}
whereby the constraint for the divergence of the magnetic field only appears as an initial condition. This preserves the volume integral of $\textbf{B}$. By contrast, the divergence of Equation~(\ref{eq:ind-dbdt5}) yields
\begin{equation}
\frac{\partial (\nabla \cdot \textbf{B})}{\partial t} + \nabla \cdot (\textbf{v} \nabla \cdot \textbf{B}) = 0.
\end{equation}
In this case, the divergence of the magnetic field is treated in the same manner as density via the continuity equation, in that the volume integral of $\nabla \cdot \textbf{B}$ is conserved. The implication is that divergence errors are advected with the fluid flow. For the previous case, divergence errors are dispersed, and it has been shown that including the term proportional to $\nabla \cdot \textbf{B}$ leads to a poorer treatment of divergence errors in SPMHD \citep{pm05}.

\subsubsection{Example: Source term in the induction equation}

The advection of a divergence `blob' illustrates the difference between solving Equations~(\ref{eq:ind-dbdt5}) and (\ref{eq:ind-dbdt6}), that is, whether the term proportional to the divergence of the magnetic field is included or not. These results mimic those presented by \citet{pm05}, with the divergence advection test originating from \citet{dedner-etal02}. The test is set up using 50$\times$50 particles on a square lattice set within $x$, $y \in [-0.5, 1.5]$. The density is uniformly $\rho=1$, with $P=6$ and $\gamma=5/3$. The initial velocity field is $\textbf{v} = [1, 1]$, such that the fluid flows from the bottom left towards the top right. The initial magnetic field is $B_x = 1 / \sqrt{4 \pi}$, with a perturbation $B_x = [(r / r_0)^8 - 2 (r / r_0)^4 + 1] / \sqrt{4 \pi}$ for $r < r_0$, with $r_0=1/\sqrt{8}$. The perturbation to the magnetic field artificially introduces a non-zero divergence to the magnetic field.

\begin{figure*}
\begin{center}
\includegraphics[width=160mm]{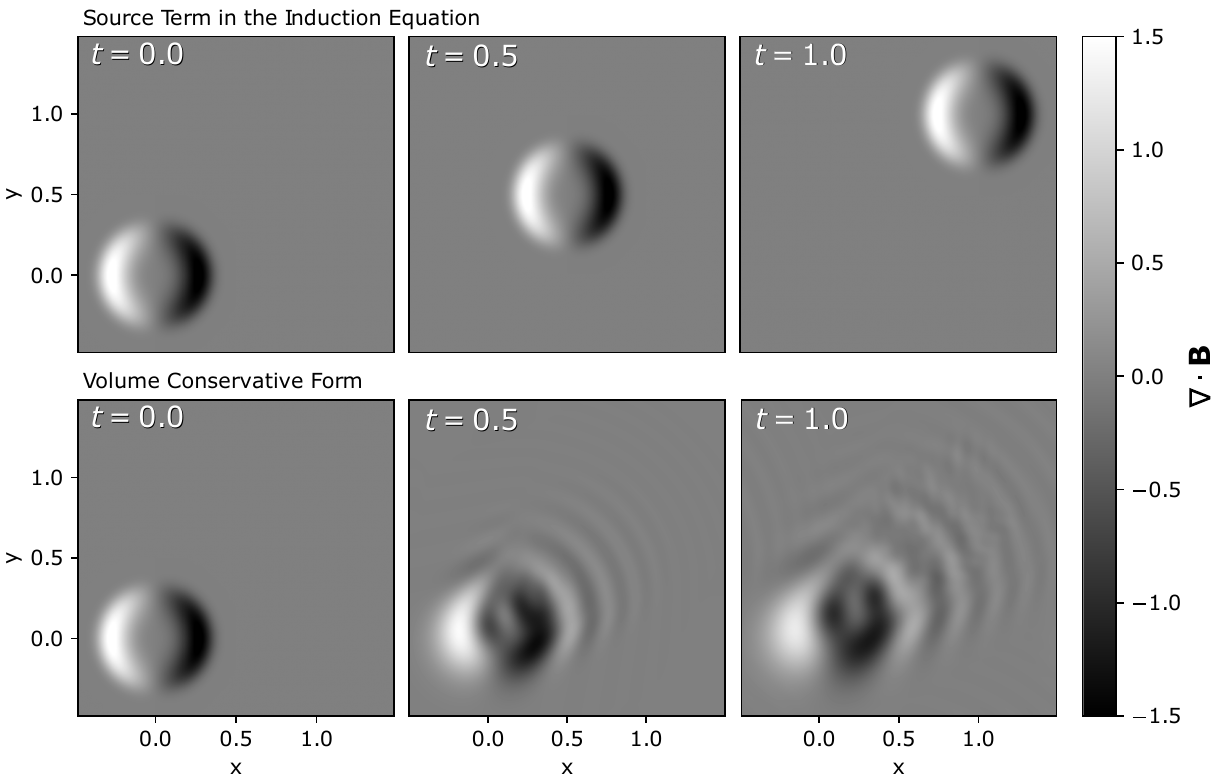}
\end{center}
\caption{The advection of $\nabla \cdot \textbf{B}$ with and without a source term related to the divergence of the magnetic field in the induction equation (top and bottom rows, respectively). With the source term (top row), the volume integral of $\nabla \cdot \textbf{B}$ is preserved and the divergence of the magnetic field is advected with the fluid flow. Without the source term (bottom row), the divergence of the magnetic field is dispersed throughout the domain.}
\label{fig:divadv}
\end{figure*}

Figure~\ref{fig:divadv} shows the transport of the divergence error between the source term (Equation~\ref{eq:ind-dbdt5}) and volume conservative (Equation~\ref{eq:ind-dbdt6}) approaches. When the term related to the divergence of the magnetic field is excluded (equivalent to the source term approach of the eight-wave solver), the divergence error is passively advected with the fluid flow. The formulation remains consistent in the presence of divergence errors. By contrast, with this term included, the volume conservative form spreads the divergence error throughout the domain.

\subsection{Artificial resistivity}
\label{sec:ar}

The discretised induction equation assumes that the magnetic field is differentiable, which would not be true at discontinuities since the magnetic field would be multi-valued. An artificial resistivity is applied to the magnetic field to smooth discontinuities over the resolution scale so that the magnetic field remains single valued. Developed by \citet{pm04a, pm05}, artificial resistivity adds dissipation to the magnetic field according to
\begin{equation}
 \frac{{\rm d}}{{\rm d}t} \left( \frac{\textbf{B}}{\rho} \right)_a = \sum_b m_b \frac{v_\text{sig,B}}{\overline{\rho}_{ab}^2} \left( \textbf{B}_a - \textbf{B}_b \right) \hat{{\bf r}}_{ab} \cdot \overline{\nabla_a W_{ab}} , \label{eq:artificialresistivity}
\end{equation}
where $\overline{\rho}_{ab} \equiv (\rho_a + \rho_b) / 2$ and $\overline{\nabla_a W_{ab}} \equiv [\nabla_a W_{ab}(h_a)/\Omega_a + \nabla_a W_{ab}(h_b)/\Omega_b] / 2$. The signal velocity, $v_{\rm sig,B}$, represents the speed of information propagation between two particles. 

Energy dissipated from the magnetic field may be added to the internal energy, $u$, through
\begin{equation}
 \frac{{\rm d}u_a}{{\rm d}t} = - \sum_b m_b \frac{v_\text{sig,B}}{\overline{\rho}_{ab}^2} \left( \textbf{B}_a - \textbf{B}_b \right)^2 \hat{{\bf r}}_{ab} \cdot \overline{\nabla_a W_{ab}} .
\end{equation}
Importantly, the deposition of dissipated magnetic energy into the internal energy is guaranteed to be positive definite. Unlike artificial viscosity, artificial resistivity is applied to both approaching and receding particles, since discontinuities in the magnetic field can occur during both compression and rarefaction, and to all components of the magnetic field (rather than just along the line of sight like artificial viscosity), since magnetic discontinuities can occur oblique to the motion \citep{pm04a, pm05}. 

One option for the signal velocity is
\begin{equation}
v_{\rm sig,B} = \frac{1}{2} \overline{\alpha_{{\rm B}, ab}} \left( v_{{\rm mhd},a} + v_{{\rm mhd},b} \right) ,
\end{equation}
which uses the averaged fast magnetosonic wave speed, $v_{\rm mhd}$. A dimensionless parameter, $\alpha_{\rm B}$, may be included per particle to switch dissipation off in regions that are not discontinuous, with $\overline{\alpha_{{\rm B},ab}} = (\alpha_{{\rm B},a} + \alpha_{{\rm B},b}) / 2$.  \citet{pm05} created a switch for $\alpha_{\rm B}$ based on analogy to the \citet{mm97} switch for artificial viscosity. In this case,
\begin{equation}
 \frac{{\rm d}\alpha_{{\rm B}, a}}{{\rm d}t} = \max\left( \frac{\vert \nabla \times \textbf{B}_a \vert}{\sqrt{\mu_0 \rho_a}} , \frac{\vert \nabla \cdot \textbf{B}_a \vert}{\sqrt{\mu_0 \rho_a}} \right) - \frac{\alpha_{{\rm B},a} - \alpha_{{\rm B}, 0}}{\tau} ,\label{eq:pm05switch}
\end{equation}
such that dissipation increases in the presence of strong gradients of the magnetic field. \citet{bkw12} found that $\alpha_{{\rm B},0} = 0$ leads to satisfactory results in their shock tube and other two-dimensional MHD tests. 

However, \citet{tp13} noted that the \citet{pm05} switch will not apply sufficient dissipation to capture discontinuities if strong shocks are present and the magnetic field is very weak (i.e., plasma $\beta \sim 10^6$--$10^{10}$), leading to numerical noise in the magnetic field and spurious growth of magnetic energy. In such a regime, using the Alfven speed instead of fast magnetosonic speed in the signal velocity is similarly problematic if shocks are present. They recommend a new switch, setting
\begin{equation}
\alpha_{{\rm B},a} = \frac{h_a \vert \nabla \textbf{B}_a \vert}{\vert \textbf{B}_a \vert} .
\end{equation}
Note that this does not require the time evolution of $\alpha_{\rm B}$. On test problems, this switch led to less overall magnetic energy dissipation than the \citet{pm05} switch. However, \citet{barnes-etal18} found that the \citet{tp13} switch was too dissipative in cosmological simulations of galaxy cluster formation.

\citet{phantom} advocated setting the signal velocity to
\begin{equation}
\label{eq:phantom-ar-switch}
v_{{\rm sig},B} = \vert \textbf{v}_{ab} \times \hat{\bf r}_{ab} \vert .
\end{equation}
No tunable dissipation parameter in the form of $\alpha_{\rm B}$ is required. It is noteworthy to realize that this switch is not dependent upon the magnetic field at all. Yet, despite this, the switch in Equation~(\ref{eq:phantom-ar-switch}) appears to provide less dissipation and work robustly in a variety of real applications, for example, within simulations of the MRI \citep{wissing-etal22} or the collapse of molecular cloud cores to stellar densities \citep{wurster-etal22}.

Though numerical stability is the purpose of including artificial resistivity in simulations, its implementation is equivalent to a (resolution-dependent) physical dissipation term $\eta \nabla^2 \textbf{B}$ with magnetic diffusivity
\begin{equation}
\eta_{\rm AR} \sim \tfrac{1}{2} v_{\rm sig,B} h .
\end{equation}
This is analogous to the translation of artificial viscosity to a physical viscosity \citep{al94, lp10}. Keep in mind that advection in SPMHD is dissipationless since it is a Lagrangian scheme. This means that, for ideal MHD calculations with only numerical dissipation, it is possible to estimate the magnetic Reynolds number or magnetic Prandtl number directly from the artificial dissipation terms \citep[e.g.,][]{wissing-etal22}.

Measurements of the magnetic Prandtl number, ${\rm Pm} = \nu / \eta$, where $\nu$ is the kinematic viscosity, typically range between ${\rm Pm} \sim 0.1$--2 when the only sources of dissipation are artificial viscosity and resistivity. There is slight dependence upon the numerical resolution as given by the particle smoothing lengths, but of more importance seems to be the specific problem under consideration. \citet{tpf16} found ${\rm Pm} \approx 1$ in simulations of supersonic, magnetized turbulence, and \citet{wissing-etal22} found ${\rm Pm} \approx 1.5$ in stratified, net-flux shearing box simulations. \citet{ws23} performed simulations of  the galactic dynamo in spiral galaxies, measuring ${\rm Pm} \approx 0.1$ in the inner radius of the galaxy and ${\rm Pm} \approx 0.7$ in the outer regions.

Overall, artificial resistivity works well to treat magnetic discontinuities. However, it is a crude tool for this purpose as it is difficult to reliably detect magnetic discontinuities, and it does not distinguish between Alfv\'{e}n waves and compressive waves. \citet{ii11} use a Godunov scheme instead of artificial resistivity to address these issues. This offers reduced numerical dissipation, but brings additional computational expense, whereas artificial resistivity is computationally inexpensive as it does not introduce additional time constraints or extra loops over the particles.

\subsubsection{Example: Brio-Wu shock tube}

\begin{figure*}
\begin{center}
\includegraphics[width=160mm]{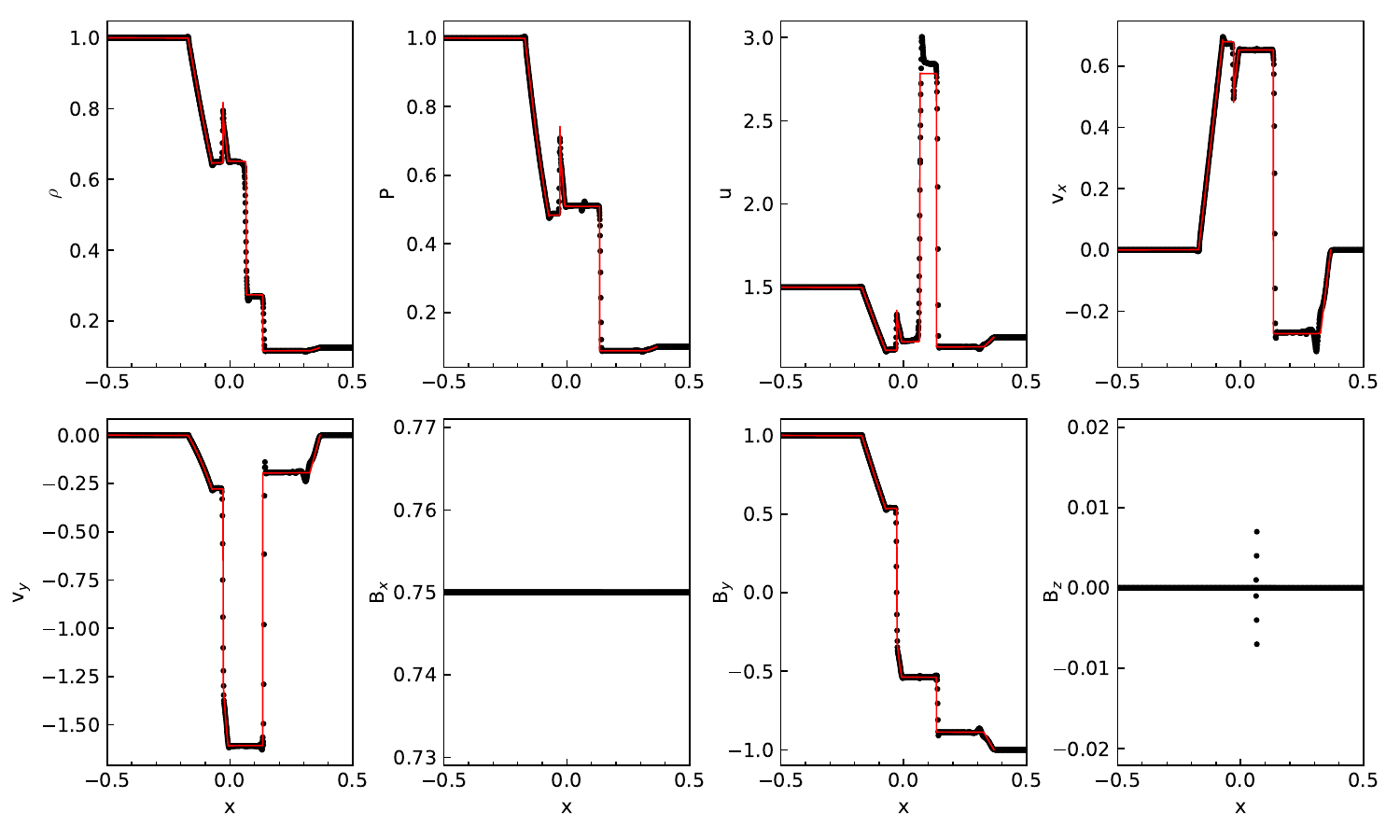}
\end{center}
\caption{Brio-Wu shocktube at $t=0.1$ with $\gamma = 5/3$. The initial configuration leads to a compound shock structure. The black circles represent the SPMHD solution, with the red line a reference solution from the \textsc{Athena} code. The magnetic field components of the SPMHD solution agree with the reference solution. Notably, $B_x$ remains constant even in the presence of density jumps and velocity discontinuities, with only small deviations in $B_x$ and $B_z$ located near the initial contact discontinuity. Results are presented using the artificial resistivity switch given by Equation~(\ref{eq:phantom-ar-switch}).\label{fig:briowu}}
\end{figure*}

The accuracy of the artificial resistivity switch given by Equation~(\ref{eq:phantom-ar-switch}) can be demonstrated with the shocktube from \citep{bw88}, which also corresponds to shocktube 5A from \citet{rj95}. This test is a magnetic extension of the \citet{sod78} shocktube. The results of this particular shocktube configuration have been presented multiple times for SPMHD, for example, amongst others, \citet{bot01, pm04a, pm04b, bkw12, tp13, hr16, tpb16, phantom, ws20}. Other shocktube configurations have been tested with SPMHD, some of which are in the preceding references.

The initial conditions are given by [$\rho$, $P$] = [1, 1] for the left state, and [$\rho$, $P$] = [0.125, 0.1] for the right state. The equation of state uses $\gamma = 5/3$, as in \citet{rj95}, which is in contrast to $\gamma=2$ in \citet{bw88}. The test is calculated in full 3D, with 800$\times$20$\times$20 particles for the left state and 400$\times$10$\times$10 particles for the right state, initially arranged on cubic lattices. The velocity is zero for both the left and right states. The magnetic field vector is [$B_x$, $B_y$, $B_z$]$_{\rm L}$ = [0.75, 1, 0] for the left state and [$B_x$, $B_y$, $B_z$]$_{\rm R}$ = [0.75, -1, 0] for the right state. There is no smoothing of the initial conditions.

Figure~\ref{fig:briowu} shows the results at $t=0.1$. The black circles represent the values on the particles, with the red line a reference solution given by the grid-based code \textsc{Athena} \citep{athena} using $10^4$ grid cells. This shocktube contains a slow compound structure, which is notably absent from the Riemann solver solution given by \citet{rj95}. SPMHD captures all of the shock structures. For the magnetic field components in particular, $B_y$ agrees with the reference solution, with the discontinuities at $x \sim -0.03$ and $x \sim 0.13$ represented over 4--8 particle spacings. In the analytic solution, the other magnetic field components remain constant over time, with $B_x = 0.75$ and $B_z = 0$. This is maintained in the SPMHD solution, except for small deviations due to the initial contact discontinuity near $x \sim 0.06$ that introduce noise into the particle arrangement. The absolute max deviations in $B_x$ and $B_z$ are on the order of $\sim 10^{-3}$.

\subsection{Euler potentials}
\label{sec:ep}

Alternative formulations to evolve the magnetic field have been investigated, though with little success. One approach that appeared promising on the surface was defining the magnetic field in terms of the Euler potentials, $\textbf{B} = \nabla \alpha_{\rm E} \times \nabla \beta_{\rm E}$, where $\alpha_{\rm E}$ and $\beta_{\rm E}$ are scalar potentials. This upholds $\nabla \cdot \textbf{B} = 0$ by construction. The scalar potentials themselves are simply advected with the fluid flow. This is trivial to implement into a Lagrangian scheme, as the scalars become just constant, passive values attached to each particle. The re-construction of the magnetic field is thus dependent solely upon the configuration of the potentials.

Calculating the gradient of each potential may be done with standard SPH derivative operators, such as the differencing derivative operator,
\begin{equation}
\nabla \alpha_{\rm E} = - \frac{1}{\Omega_a \rho_a} \sum_b m_b \left( \alpha_{{\rm E}, a} - \alpha_{{\rm E}, b} \right) \nabla_a W_{ab}(h_a) ,
\end{equation}
\begin{equation}
\nabla \beta_{\rm E} = - \frac{1}{\Omega_a \rho_a} \sum_b m_b \left( \beta_{{\rm E}, a} - \beta_{{\rm E}, b} \right) \nabla_a W_{ab}(h_a) .
\end{equation}

Plugging the re-constructed magnetic field into the conservative, Lagrangian equations of motion (see section~\ref{sec:accelerations}) yields a numerically stable scheme, and numerical formulations of this type have been used in several astrophysical applications, such as star formation \citep{pb07, pb08, pb09}, neutron star mergers \citep{pr06} and the magnetic fields of galaxies \citep{dp08, kotarba-etal09}.

However, the Euler potentials are incapable of correctly capturing the full range of magnetic field topologies \citep{brandenburg10}. Since the potentials are passively advected, there is a one-to-one mapping from the initial magnetic field topology to any evolved field. This is problematic for rotating discs, for example, as the magnetic field would be essentially reset with each turn once the particles return to their initial positions. Even worse is that the Euler potentials are incapable of modelling dynamos, and are incompatible with linear diffusion operators \citep{brandenburg10}.

Star formation calculations highlight the preceding limitations. SPMHD calculations that evolve the magnetic field via the induction equation produce magnetically-driven jets and outflows \citep[e.g.,][]{burzle-etal11b, ptb12, btp14, tsukamoto-etal15a, lbp15, hr16, wbp18a, tsukamoto-etal18, ws20, garcia-senz-etal23}. However, no such outflows are produced when similar calculations are performed using the Euler potentials \citep{pb07}. At present, it seems there is little use for the Euler potentials in astrophysical simulation, except perhaps as a secondary check for numerical artefacts \citep{ds09}.

\subsection{Vector potential}
\label{sec:vecp}

Defining the magnetic field in terms of the vector potential, $\textbf{B} = \nabla \times {\bf A}$, would seem an ideal choice. Doing so would guarantee a divergence-free magnetic field, without any of the topological restrictions associated with the Euler potentials. However, SPMHD implementations of the vector potential appear to be plagued with numerical instabilities \citep{price10}.

An evolution equation for the vector potential can be obtained by combining $\textbf{B} = \nabla \times {\bf A}$ with the induction equation (i.e., Equation~\ref{eq:ind-partial}), which yields
\begin{equation}
\frac{{\rm d}{\bf A}}{{\rm d}t} = \textbf{v} \times \left( \nabla \times {\bf A} \right) + \left( \textbf{v} \cdot \nabla \right) {\bf A} + \nabla \phi ,
\end{equation}
where $\nabla \phi$ represents the gauge choice, arising as a constant of integration. \citet{price10} used the gauge $\phi = - \textbf{v} \cdot {\bf A}$, yielding
\begin{equation}
\frac{{\rm d}{\bf A}}{{\rm d}t} = - {\bf A} \times \left( \nabla \times \textbf{v} \right) - \left( {\bf A} \cdot \nabla \right) \textbf{v} .
\end{equation}
This has the effect of moving the derivative onto $\textbf{v}$ instead of ${\bf A}$, which makes the discretisation Galilean invariant. In spite of this, using the same approach to calculate the Lorentz force as employed with the Euler potentials, whereby the re-constructed magnetic field is used within the equations of motion (see section~\ref{sec:accelerations}), will lead to numerical instability and runaway energy growth \citep{price10}. That this occurs for the vector potential and not the Euler potentials is understandable because the Euler potentials are held constant. 

Numerical instability and runaway energy growth occur due to the mismatch of variables between the evolution of the magnetic energy (evolving ${\bf A}$) and kinetic energy (equations of motion using $\textbf{B}$), such that the total energy in this hybrid formulism is not conserved. In theory, this is solved by deriving the consistent, conservative equations of motion from the Lagrangian that directly use the vector potential. Unfortunately, this only seems to lead to a different set of numerical issues. The resultant equations of motion contain three terms, with the first term acting like a negative magnetic pressure, the second term involving second derivatives of the kernel (which are known to be highly sensitive to particle disorder), and the third term nested low-order derivative estimates \citep{price10}. These equations do fundamentally represent the MHD equations of motion in the continuum limit, such that, for example, the second and third terms partially encode the magnetic pressure to offset the negative pressure of the first term, but the inherent numerical issues mean that this is not realized in practice.

\citet{se15} explored options to revive the hybrid formulism, such as smoothing the re-constructed magnetic field, adding numerical dissipation of the vector potential, and choosing the Coulomb gauge, $\phi = \nabla \cdot {\bf A}$, which is upheld through mixed hyperbolic / parabolic divergence cleaning (more on this in section~\ref{sec:cleaning}). \citet{tu-etal22} also smoothed the re-constructed magnetic field, with the Weyl gauge, $\phi = 0$, albeit within a meshless finite mass (MFM) scheme. Further testing is required on the robustness of these approaches. \citet{tp23} derived a novel discretisation for the evolution equation of ${\bf A}$ in terms of a volume integral, but which also failed to produce a valid numerical solution. At present, it remains to be demonstrated that there is a formulation of the vector potential in SPMHD that is viable for general astrophysical application.

\section{Accelerations from the magnetic field}
\label{sec:accelerations}

The equations of motion for SPMHD are derived from the Lagrangian. The implication of doing this is that the discretised equations of motion are the physical equations governing the system of discrete particles. This imparts a number of desirable properties, such as exact conservation of momentum, energy and entropy. Some early work directly discretised the Lorentz force, ${\bf J} \times \textbf{B}$, where ${\bf J} = \nabla \times \textbf{B} / \mu_0$ is the current density, but these formulations do not provide the conservation properties of the Lagrangian approach \citep{gm77, mwd95} and are poor at capturing shocks \citep{morris96thesis}.

\subsection{Conservative equations of motion}

The SPMHD Lagrangian is
\begin{equation}
L_{\rm SPMHD} = \sum_a m_a \left( \frac{v_a^2}{2}  - u_a - \frac{B_a^2}{2 \mu_0 \rho_a} \right) .
\end{equation}
The SPMHD equations of motion can be derived by using a variational approach \citep{pm04b, price12}. The only ingredients necessary are the prescription for density (via the density summation in Equation~\ref{eq:dens-sum}) and the specification on how the magnetic field is evolved. The induction equation given by Equation~(\ref{eq:spmhd-ind-brho}) is assumed (or equivalently Equation~\ref{eq:spmhd-ind-b}).

The resultant equations of motion from the Lagrangian are
\begin{align}
\label{eq:spmhd-momeq}
 \frac{{\rm d}\textbf{v}_a}{{\rm d}t} = &- \sum_b m_b \bigg[ \frac{P_a}{\Omega_a \rho_a^2} \nabla_a W_{ab}(h_a) \nonumber \\
 & \hspace{1cm} + \frac{P_b}{\Omega_b \rho_b^2} \nabla_a W_{ab}(h_b) \bigg] \nonumber \\
& - \frac{1}{2\mu_0} \sum_b m_b \bigg[ \frac{B_a^2}{\Omega_a \rho_a^2} \nabla_a W_{ab}(h_a) \nonumber \\
& \hspace{1cm} + \frac{B_b^2}{\Omega_b \rho_b^2} \nabla_a W_{ab}(h_b) \bigg] \nonumber \\
& + \frac{1}{\mu_0} \sum_b m_b \bigg[ \frac{\textbf{B}_a}{\Omega_a \rho_a^2} \textbf{B}_a \cdot \nabla_a W_{ab}(h_a) \nonumber \\
& \hspace{1cm} + \frac{\textbf{B}_b}{\Omega_b \rho_b^2} \textbf{B}_b \cdot \nabla_a W_{ab}(h_b) \bigg] . 
\end{align}
The second term in Equation~(\ref{eq:spmhd-momeq}) represents an isotropic magnetic pressure and the third term an anistropic magnetic tension. The correspondence between the SPMHD momentum equation and the stress tensor is clear upon examination of Equation~(\ref{eq:spmhd-momeq}) with Equations~(\ref{eq:mhdmomentumeqn}) and (\ref{eq:stresstensor}), and, indeed, Equation~(\ref{eq:spmhd-momeq}) could be written in terms of the stress tensor as
\begin{align}
 \frac{{\rm d}v_a^i}{{\rm d}t} = \sum_b m_b \Bigg[& \frac{S^{ij}_a}{\Omega_a \rho_a^2} \frac{\partial W_{ab}(h_a)}{\partial x_a^j} \nonumber \\
 & + \frac{S^{ij}_b}{\Omega_b \rho_b^2} \frac{\partial W_{ab}(h_b)}{\partial x_a^j} \Bigg] .
 \end{align}
 
Note, however, that while these equations of motion exactly conserve momentum, energy and entropy (provided they are used in conjunction with the density summation and the induction equation prescribed in their derivation), they do not conserve angular momentum. The anisotropic magnetic tension is not invariant to rotation. Notably, this term is derived solely from the numerical choice for the induction equation.

\cite{ws20} investigated switching the gradient operator in the momentum equation to one based on geometric density averaging, given by
\begin{equation}
 \frac{{\rm d}v_a^i}{{\rm d}t} = \sum_b m_b \frac{S^{ij}_a + S^{ij}_b}{\rho_a \rho_b} \overline{\nabla^j_a W_{ab}} .
 \end{equation}
 This type of gradient operator is used for the (thermodynamic) pressure gradient in the {\sc Gasoline2} code \citep{gasoline2}, and promises to reduce artificial surface tension type effects \citep{agertz-etal07}. 
For SPMHD, \citet{ws20} find that the geometric density average formulism has a lower numerical resolution requirement to produce magnetically-driven outflows from the gravitational collapse of magnetised molecular cloud cores. This is promising and warrants further examination.
 
Writing the momentum equation in terms of the stress tensor creates a complication in that the anisotropic tension contains a component due to monopole moments. The third term in Equation~(\ref{eq:spmhd-momeq}) is equivalent to $(\textbf{B} \cdot \nabla) \textbf{B} / \mu_0 \rho + (\nabla \cdot \textbf{B}) \textbf{B} / \mu_0 \rho$ in the continuum limit. The force contributions proportional to $\nabla \cdot \textbf{B}$ are present in order to be momentum conserving in the presence of monopoles. The issue of monopole forces in SPMHD is complicated, in that even for a magnetic field that is constant and uniform (i.e., $\nabla \cdot \textbf{B} = 0$), the discretisation used in the momentum equation may produce monopole forces. This is related to the `E0' errors often discussed in relation to the pressure gradient \citep{rha10, mlp12, da12, hopkins15}.

The real challenge with monopole accelerations is not related to the discretisation, however, but arise from any non-zero divergence errors that may be present in the magnetic field. The induction equation in SPMHD makes no guarantee of a divergence-free magnetic field. The detrimental effect of $(\nabla \cdot \textbf{B}) \textbf{B} / \mu_0 \rho$ is most strongly connected to the severity of non-zero divergence errors. 

Compounding the considerations of monopole accelerations is that the conservative equations of motion themselves are tensile unstable. The anisotropic magnetic tension is an attractive force, and if it exceeds the isotropic pressure, then the particles will unphysically clump. For Equation~(\ref{eq:spmhd-momeq}), this occurs for plasma $\beta < 1$ \citep{pm85, bot01, bot04}. Importantly, this criterion is only based upon the magnitude of the magnetic field. Fortunately, removing the tensile instability is not difficult.

\subsection{Removing the tensile instability}

The approach that seems to work best to counteract the tensile instability is to explicitly subtract the non-physical force arising from the monopole contribution \citep{bot01}. This aligns with the source-term approach of the eight-wave solver \citep{powell94, powell-etal99}. Importantly, the removal of monopole accelerations must use the same discretisation for $\nabla \cdot \textbf{B}$ as in the momentum equation, that is,
\begin{align}
\label{eq:divb-subtract}
\left( \frac{{\rm d}{\textbf{v}_a}}{{\rm d}t} \right)_{\rm divB} = & - \hat{\bm{\beta}}_a \sum_b m_b \Bigg[ \frac{\textbf{B}_a}{\Omega_a \rho_a^2} \cdot \nabla_a W_{ab}(h_a)\nonumber \\
& \hspace{8mm} + \frac{\textbf{B}_b}{\Omega_b \rho_b^2} \cdot \nabla_a W_{ab}(h_b) \Bigg] ,
\end{align}
where $\hat{\bm{\beta}}_a$ used to regulate the strength of the applied correction, with $\hat{\bm{\beta}}_a = \textbf{B}_a$ corresponding to a full correction of the force term. Adding Equation~(\ref{eq:divb-subtract}) to the conservative momentum equation (Equation~\ref{eq:spmhd-momeq}) yields a numerically stable solution. 

The consequence of removing monopole accele\-rations is that the conservative equations of momentum no longer conserve momentum. The severity of this non-conservation of momentum is dependent upon the magnitude of divergence errors. In the worst case, this can corrupt the obtained solution \citep[for a dramatic example of this, see][]{tp12}. Thus, controlling the magnitude and growth of divergence errors will improve momentum conservation.

\citet{bot04} showed that the tensile instability can be corrected with $\hat{\bm{\beta}} = \tfrac{1}{2} \textbf{B}$, thereby subtracting only $\tfrac{1}{2} (\nabla \cdot \textbf{B}) \textbf{B} / \mu_0 \rho$. In theory, this provides a factor of two improvement on the non-conservation of momentum. \citet{bkw12} recommended this for general SPMHD calculations, though \citet{tp12} found that this might cause numerical artefacts as, while technically sufficient to prevent instability, it may leave particles in a near-pressureless state. Compounding this is that subtracting the full strength term {($\hat{\bm{\beta}}_a = \textbf{B}_a$) causes dispersive errors in slow magnetosonic waves. The half strength term ($\hat{\bm{\beta}}_a = \tfrac{1}{2} \textbf{B}_a$) does not introduce these type of errors \citep[see section~\ref{sec:mhdwave}, c.f.][]{iwasaki15}. 

Since the tensile instability only manifests for plasma $\beta < 1$, \citet{bot06}, \citet{phantom} and \citet{ws20} have proposed schemes where the tensile instability correction is switched off in pressure-dominated regimes, suggesting, respectively,
\begin{equation}
\hat{\bm{\beta}}_a =
\begin{cases}
\textbf{B}_a & \beta < 2 \\
\textbf{B}_a (10 - \beta) / 8 & 2 < \beta < 10 \\
0 & \beta \geq 10
\end{cases}
\end{equation}
and
\begin{equation}
\hat{\bm{\beta}}_a =
\begin{cases}
\textbf{B}_a & \beta < 1 \\
\textbf{B}_a (2 - \beta) & 1 < \beta < 2 \\
0 & \beta \geq 2
\end{cases}
\end{equation}
Both provide a linear transition from full correction to no correction once plasma $\beta$ is above a critical threshold. This provides exact momentum conservation when possible.

An alternative approach to deal with the tensile instability is to use a more accurate derivative estimate for the anisotropic term \citet{morris96thesis}. This uses the conservative form for the isotropic hydrodynamic and magnetic pressure (first and second terms in Equation~\ref{eq:spmhd-momeq}), but replaces the anisotropic force (third term in Equation~\ref{eq:spmhd-momeq}) with
\begin{equation}
\label{eq:spmhd-morris}
\left( \frac{{\rm d} v_a^i}{{\rm d}t} \right)_{\rm aniso} = \frac{1}{\mu_0} \sum_b m_b \frac{B_b^i B_b^j - B_a^i B_a^j}{\rho_a \rho_b} \overline{\nabla_a^j W_{ab}(h_a)} .
\end{equation}
This is not momentum conserving, but yields a numerically stable solution. A disadvantage is that the \citet{morris96thesis} approach cannot be switched off, and also has dispersive errors in slow magnetosonic waves \citep{iwasaki15}.

\subsubsection{Example: Propagation of an isolated wave}
\label{sec:mhdwave}

The tensile instability correction term, depending upon its implementation, can introduce dispersive errors to slow MHD waves. Here, this is demonstrated using the test from \citet{iwasaki15} of the propagation of an isolated wave. The initial density is $\rho = 1$, with 320$\times$160 particles arranged on a square lattice in the domain $x \in [-2, 2]$ and $y \in [-1, 1]$. An isothermal equation of state is used, with $P=1$ and sound speed, $c_{\rm s}=1$. The magnetic field is uniform along the $x$-direction with plasma $\beta = 0.1$. The test is performed using the cubic and quintic spline kernels.

Figure~\ref{fig:mhdwave} shows the rightward propagating wave at $t=0.5$ for $\hat{\bm{\beta}} = \textbf{B}$ (top panel) and $\hat{\bm{\beta}} = \tfrac{1}{2} \textbf{B}$ (bottom panel), where $\hat{\bm{\beta}}$ is the leading term in the tensile instability correction term given by Equation~(\ref{eq:divb-subtract}). The location of the expected peak of the wave at $t=0.5$ is given by the vertical dashed line ($x = 0.5$). For $\hat{\bm{\beta}} = \textbf{B}$, dispersive errors cause supersonic velocities preceding the wave. In this case, the quality of the kernel affects the result, with the cubic spline showing a larger phase velocity than the quintic spline. For  $\hat{\bm{\beta}} = \tfrac{1}{2} \textbf{B}$, there are no dispersive errors and the wave propagates (mostly) as expected. 

\begin{figure}
\begin{center}
\includegraphics[width=85mm]{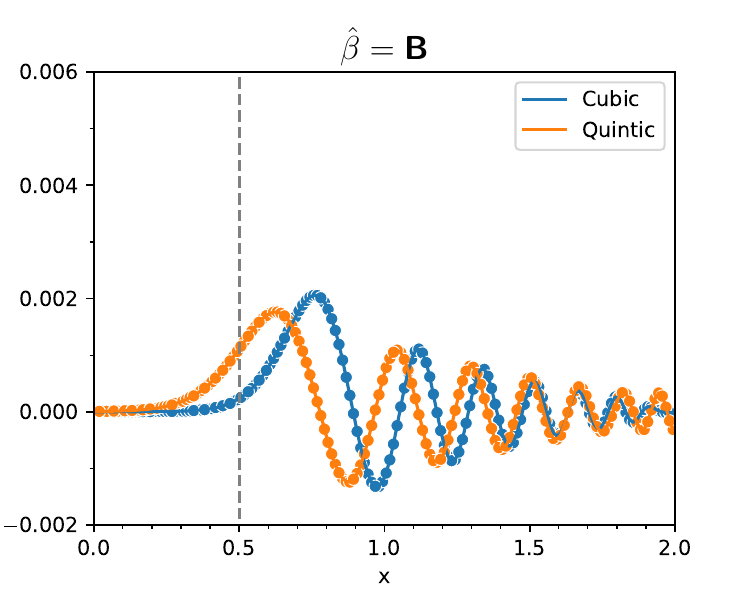}
\includegraphics[width=85mm]{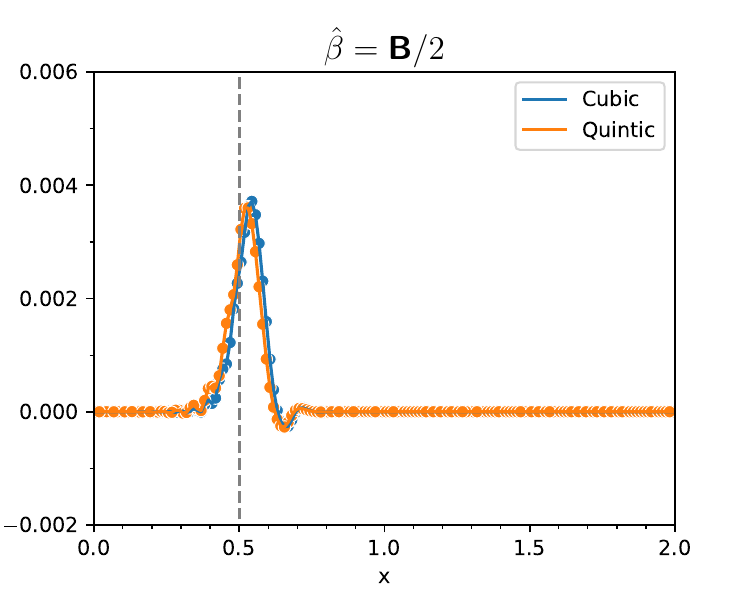}
\end{center}
\caption{The propagation of an isolated wave in a strongly magnetized medium (plasma $\beta = 0.1$) with the full tensile instability correction term (top) and the correction term reduced by half (bottom), that is, $\hat{\bm{\beta}} = \textbf{B}$ and $\hat{\bm{\beta}} = \tfrac{1}{2} \textbf{B}$ as used in Equation~(\ref{eq:divb-subtract}). Applying the full correction term leads to supersonic dispersive errors, whereas the half correction term yields results consistent with the expected solution.}
\label{fig:mhdwave}
\end{figure}

\section{Divergence-free constraint of the magnetic field}
\label{sec:divb}

Upholding the divergence-free constraint on the magnetic field is critical. In one respect, real magnetic fields are purely solenoidal, so modelling magnetic fields that preserve this topological constraint is only logical. From a theoretical perspective, magnetic fields within simulation should avoid unphysical configurations. In another respect, there are numerical justifications for minimizing divergence errors. Doing so avoids numerical artefacts in the momentum equation and the overall conservation of momentum and energy. 

An important question is: how does one define a ``divergence-free magnetic field'' numerically? One such definition has already been used in Equation~(\ref{eq:divb-subtract}), whereby
\begin{align}
\label{eq:divb-symmetric}
\nabla \cdot \textbf{B}_a = & \rho_a \sum_b m_b \bigg[ \frac{\textbf{B}_a}{\Omega_a \rho_a^2} \cdot \nabla_a W_{ab}(h_a) \nonumber \\
& \hspace{8mm}+ \frac{\textbf{B}_b}{\Omega_b \rho_b^2} \cdot \nabla_a W_{ab}(h_b) \bigg] .
\end{align}
Instead of this symmetric gradient estimate, one could instead measure the divergence of the magnetic field with a difference gradient estimate according to
\begin{equation}
\label{eq:divb-difference}
\nabla \cdot \textbf{B}_a = - \frac{1}{\Omega_a \rho_a} \sum_b m_b (\textbf{B}_a - \textbf{B}_b) \cdot \nabla_a W_{ab}(h_a) .
\end{equation}
Measuring zero divergence of the magnetic field in one metric does not guarantee zero in another. As a case in point, neither of the above discretisations are zero when the magnetic field is specified in terms of the Euler potentials, even though it intrinsically guarantees a divergence-free magnetic field \citep{tp12}. Nonetheless, minimizing divergence error in one discretisation will often reduce divergence error in another discretisation. This is important for SPMHD, as it means the detrimental side effects of the tensile instability correction will benefit from divergence control in a different discretisation.

At present, there are no robust methods in SPMHD for exactly preserving the divergence-free constraint on the magnetic field. The challenges faced with the Euler potentials and vector potential implementations are discussed in sections~\ref{sec:ep} and \ref{sec:vecp}. Instead, it would appear that the best option currently is to only approximately uphold this constraint.

\subsection{Hyperbolic / parabolic divergence cleaning}
\label{sec:cleaning}

\citet{dedner-etal02} introduced a divergence cleaning approach that transports divergence errors away from their source and damps them through a set of coupled hyperbolic and parabolic equations. In essence, divergence errors are propagated through a damped wave equation. A new, scalar field, $\psi$, is coupled to the magnetic field to facilitate this. 

\citet{tp12} adapted the \citet{dedner-etal02} divergence cleaning scheme for SPMHD, with later improvements by \citet{tpb16}. The cleaning equations, in continuum form, are given by
\begin{equation}
\label{eq:cleaning-continuum1}
\frac{{\rm d}\textbf{B}}{{\rm d}t} = - \nabla \psi ,
\end{equation}
\begin{equation}
\label{eq:cleaning-continuum2}
\frac{{\rm d}}{{\rm d}t} \left( \frac{\psi}{c_{\rm h}} \right) = - c_{\rm h} (\nabla \cdot \textbf{B}) - \frac{1}{\tau} \left( \frac{\psi}{c_{\rm h}} \right) - \frac{1}{2} \left( \frac{\psi}{c_{\rm h}} \right) (\nabla \cdot \textbf{v}) .
\end{equation}
This formulation differs from \citet{dedner-etal02}, in that Equations~(\ref{eq:cleaning-continuum1}) and (\ref{eq:cleaning-continuum2}) use Lagrangian derivatives, use $\psi / c_{\rm h}$ for the evolved variable, and account for compression and rarefaction of the fluid. Also accounted for are wave cleaning speed, $c_{\rm h}$, and parabolic damping, $\tau$, that vary in time. \citet{tpb16} showed that the above system of equations can be combined to create a generalised wave equation of the form
\begin{align}
& \frac{{\rm d}}{{\rm d}t} \left[ \frac{1}{\sqrt{\rho} c_{\rm h}} \frac{{\rm d}}{{\rm d}t} \left( \frac{\psi}{\sqrt{\rho}c_{\rm h}} \right) \right] - \frac{\nabla^2 \psi}{\rho} \nonumber \\
& \hspace{15mm} + \frac{{\rm d}}{{\rm d}t} \left[ \frac{1}{\sqrt{\rho}c_{\rm h}} \left( \frac{\psi}{\sqrt{\rho}c_{\rm h} \tau} \right) \right] = 0.
\end{align}
If $c_{\rm h}$, $\tau$, $\rho$ and the fluid velocity are held constant, this reduces to the usual damped wave equation, specified in the original \citet{dedner-etal02} formulation,
\begin{equation}
\frac{\partial^2 \psi}{\partial t^2} - c_{\rm h}^2 \nabla^2 \psi + \frac{1}{\tau} \frac{\partial \psi}{\partial t} = 0.
\end{equation}

The SPMHD discretised cleaning equations are
\begin{align}
\frac{{\rm d}}{{\rm d}t} \left( \frac{\textbf{B}}{\rho} \right)_a = &- \sum_b m_b \bigg[ \frac{\psi_a}{\Omega_a \rho_a^2} \nabla_a W_{ab}(h_a) \nonumber \\
& \hspace{10mm} + \frac{\psi_b}{\Omega_b \rho_b^2} \nabla_a W_{ab}(h_b) \bigg] ,
\end{align}
\begin{align}
\frac{{\rm d}}{{\rm d}t} \left( \frac{\psi}{c_{\rm h}} \right)_a = & \frac{c_{{\rm h},a}}{\Omega_a \rho_a} \sum_b m_b \left( \textbf{B}_a - \textbf{B}_b \right) \cdot \nabla_a W_{ab}(h_a) \nonumber \\
& - \frac{1}{\tau} \left( \frac{\psi}{c_{\rm h}} \right)_a \nonumber \\
& + \frac{1}{2} \left( \frac{\psi}{c_{\rm h}} \right)_a \sum_b m_b \textbf{v}_{ab} \cdot \nabla_a W_{ab}(h_a) .
\end{align}
Note that the difference estimate is used for $\nabla \cdot \textbf{B}$ (Equation~\ref{eq:divb-difference}). These `constrained' hyperbolic / parabolic divergence cleaning equations have formal guarantees about numerical stability. The discretised cleaning equations were derived by considering the energy content of the $\psi$ field, $e_\psi = \psi^2 / 2 \mu_0 \rho c_{\rm h}^2$, ensuring that any change in the magnetic energy is conserved by an equal but opposite change in $\psi$ energy so that total energy is conserved. Imposing this constraint further ensures that the parabolic damping will only ever siphon energy from the magnetic field.

The hyperbolic wave speed, $c_{\rm h}$, is typically taken to be the fast magnetosonic wave speed, so that propagation is at the fastest rate permissible within the Courant timestep condition. This is given, per particle, by $\Delta t_a = C h_a / c_{{\rm h},a}$ where $C \sim 0.3$ is the Courant factor. Since the hydrodynamical timestep is already constrained by the fast magnetosonic speed, this does not impose any additional timestep constraint. It is possible to use a faster speed, so long as the timestep is commensurately decreased \citep{dw21}. The rate of damping is set by
\begin{equation}
\tau = \frac{h}{\sigma c_{\rm h}} ,
\end{equation}
with $\sigma = 1$ an optimal choice \citep{tp12, barnes-etal18}.

\begin{figure*}
\begin{center}
\includegraphics[width=160mm]{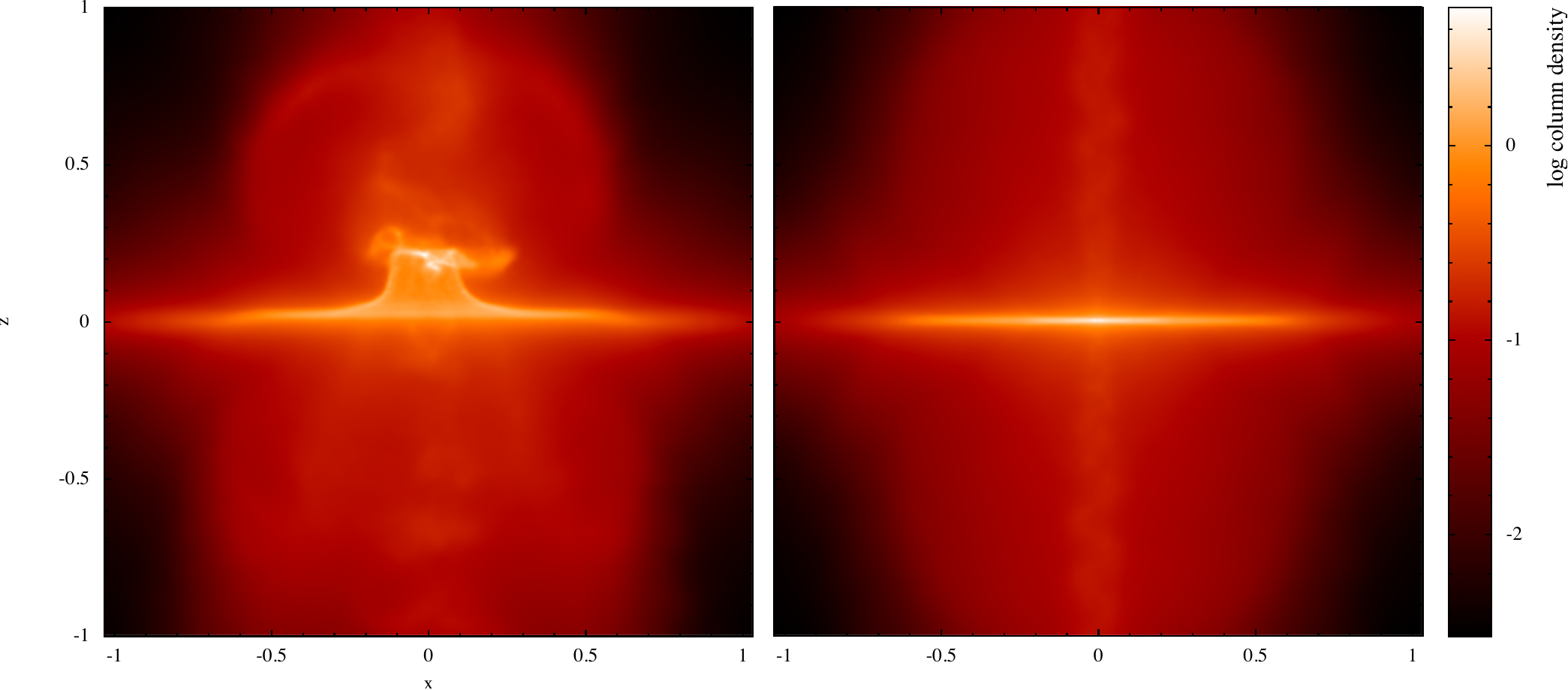}
\end{center}
\caption{The gravitational collapse of a molecular cloud core. Without divergence control (left panel), divergence errors lead to material becoming kicked out of the plane of the disc due to momentum errors arising from the tensile instability correction. With divergence cleaning (right panel), divergence errors are kept sufficiently low to avoid catastrophic momentum injection, and the disc remains planar. Reproduced from Figure 21 in \citet{tp12}.}
\label{fig:jet}
\end{figure*}

The divergence error within a calculation can be measured by the dimensionless quantity
\begin{equation}
\epsilon_{\rm divB} = \frac{h \vert \nabla \cdot \textbf{B} \vert}{\vert \textbf{B} \vert} .
\end{equation}
Typically, `constrained' divergence cleaning keeps the mean $\epsilon_{\rm divB}$ in a simulation around ${\sim}1$\%. This is sufficient for practical applications, and, generally, reduces average divergence error by about an order of magnitude for most test problems and astrophysical applications. Divergence cleaning can subsequently yield multiple orders of magnitude improvement in momentum conservation through the connection of the tensile instability correction with the divergence of the magnetic field. An example of this is shown in Figure~\ref{fig:jet}, which showcases the gravitational collapse of a molecular cloud core from \citet{tp12} (their Figure~21). Without divergence cleaning, material is ejected out of the plane of the disc due to spurious momentum gain from the tensile instability correction arising from divergence errors. With divergence cleaning, the disc remains stable over long-term evolution. Figure~\ref{fig:jet-mom} shows the corresponding total momentum change. A sink particle is introduced around one free-fall time. Divergence cleaning improves momentum conservation by approximately two orders of magnitude, which is sufficient to keep the sink particle within the disc.

\begin{figure}
\begin{center}
\includegraphics[width=75mm]{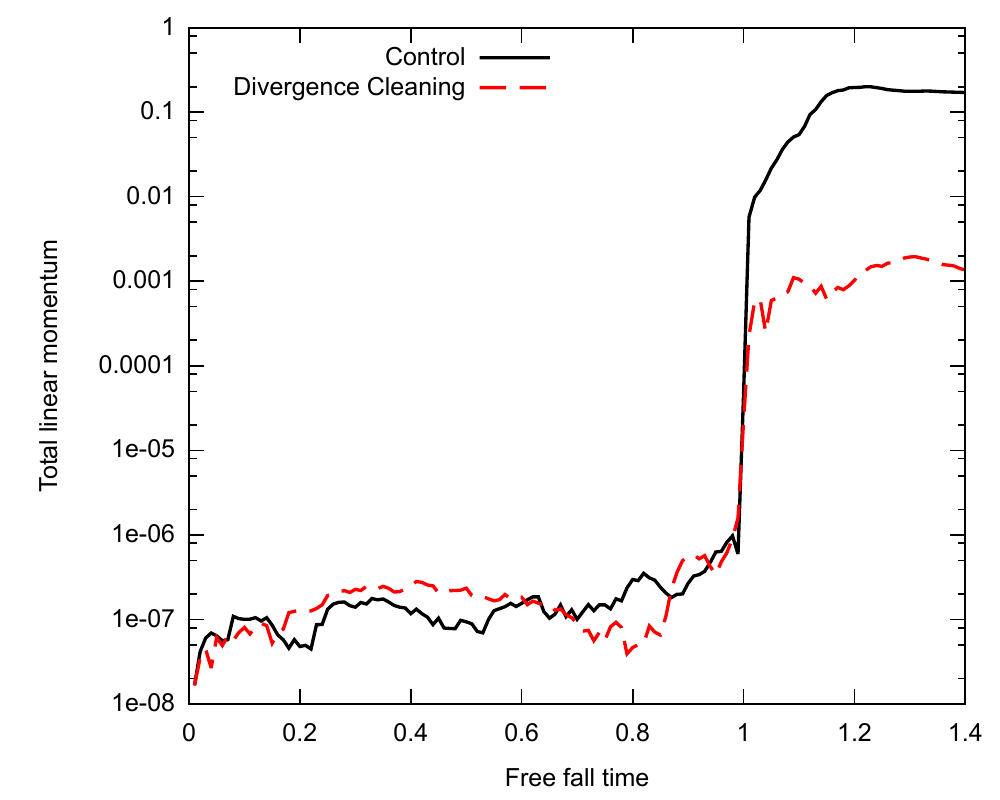}
\end{center}
\caption{Change in total linear momentum over 1.4 free-fall times for the gravitational collapse of a molecular cloud core (see Figure~\ref{fig:jet}). A sharp injection of momentum occurs when a sink particle is introduced around one free-fall time. Divergence cleaning improves the conservation of momentum by two orders of magnitude, which is sufficient to keep the sink particle within the disc. Reproduced from Figure 22 in \citet{tp12}.}
\label{fig:jet-mom}
\end{figure}

Importantly, the `constrained' divergence cleaning scheme is stable in the presence of free boundaries or sharp density contrasts \citep{tp12}. \citet{pm05} tested a non-conservative scheme on the Orszag-Tang vortex \citep{ot79}, finding that it gives, at best, a factor two reduction in average divergence error, but could potentially increase overall divergence error through spurious energy creation (a `divergence creating' scheme). \citet{sdb13} demonstrate that a non-conservative scheme can corrupt shocktube solutions with sharp density jumps.

Although divergence cleaning uses the difference derivative estimate for $\nabla \cdot \textbf{B}$, as given by Equation~(\ref{eq:divb-difference}), it is possible in principle to use other discretisations. As an example, \citet{tp12} constructed `constrained' divergence cleaning equations that used the symmetric derivative estimate (Equation~\ref{eq:divb-symmetric}). Though numerically stable, this low-order estimate was found to be overly dissipative and could introduce artefacts into the physical portions of the magnetic field since this estimate is sensitive to particle disorder. 

\subsubsection{Example: 3D Orszag-Tang vortex}

The Orszag-Tang vortex \citep{ot79} is a widely used test that consists of interacting vortices that incite magnetized turbulence, and is used here to demonstrate the efficacy of mixed hyperbolic/parabolic divergence cleaning. Results for this test have been shown for SPMHD many times, for example, \citet{pm05, bot06, rp07, ds09, bkw12, tp12, tp13, phantom, ws20}. Originally a 2D test, it is extended to 3D for this test by adding a small thickness in $z$. The simulation domain is $x$, $y \in [0, 1]$ and $z \in [0, 3/128]$, with 512$\times$512$\times$12 particles arranged on a cubic lattice. The initial conditions are $\rho$ = 25/(36 $\pi$), $P$ = 5/(12$\pi$), $\textbf{v}$ = [-$\sin(2 \pi y)$, $\sin(2 \pi x)$, 0] and $\textbf{B}$ = [-$\sin(2 \pi y)$, $\sin(4 \pi x)$, 0] with $\gamma$ = 5/3. The calculations use the artificial resistivity switch given by Equation~(\ref{eq:phantom-ar-switch}).

Figure~\ref{fig:ot} shows the density cross-sections at $t=0.5$ and $t=1$ for the calculation with divergence cleaning. The divergence error can be measured by the dimensionless quantity $\epsilon_{\rm divB} = h \vert \nabla \cdot \textbf{B} \vert / \vert \textbf{B} \vert$, which provides a standard metric for comparison. Figure~\ref{fig:ot-hdivbb} shows the median $\epsilon_{\rm divB}$ over time for calculations of the 3D Orszag-Tang vortex with and without divergence cleaning. The shaded regions represent the inter-quartile range, that is, the 25$^{\rm th}$ to 75$^{\rm th}$ percentiles. There is a significant reduction in divergence error when divergence cleaning is applied. The median $\epsilon_{\rm divB}$ with divergence cleaning is $\sim 0.1\%$ (the average $\epsilon_{\rm divB}$ is also $\sim 0.1\%$). This is below the 25$^{\rm th}$ percentile without divergence cleaning.

\begin{figure}
\includegraphics[width=85mm]{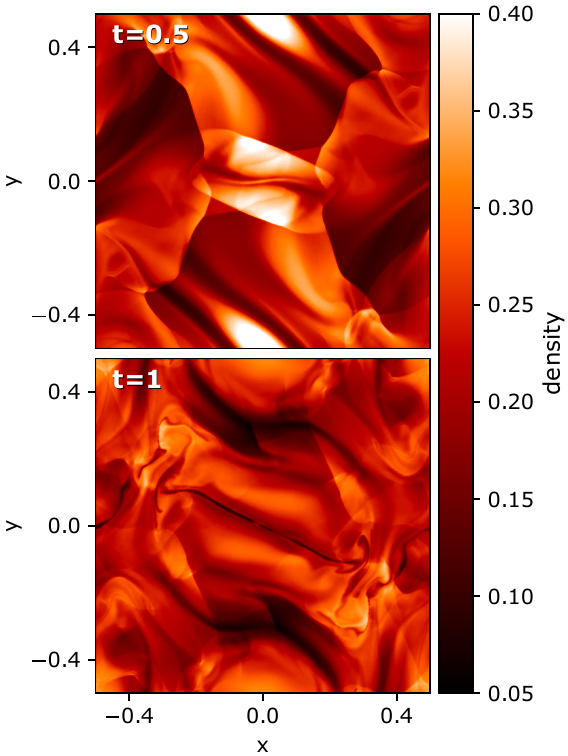}
\caption{Density cross-sections of the 3D Orszag-Tang vortex.}
\label{fig:ot}
\end{figure}

\begin{figure}
\includegraphics[width=85mm]{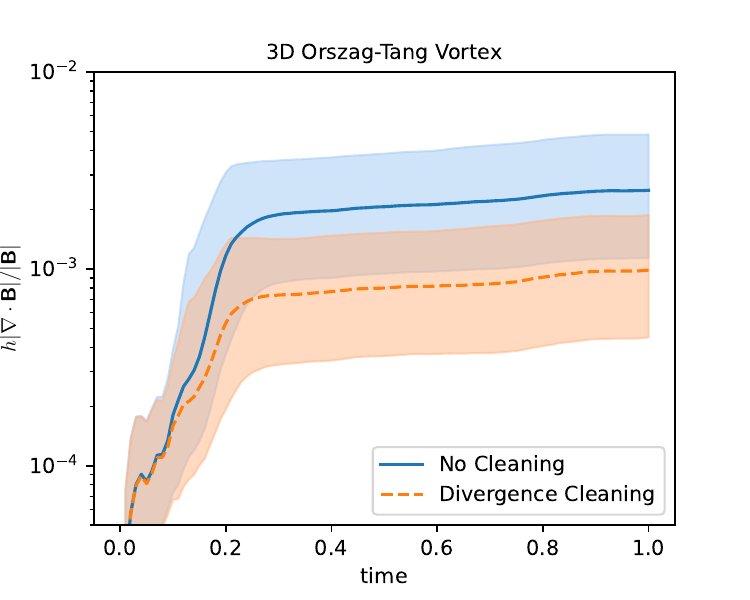}
\caption{The median divergence error across all particles for the 3D Orszag-Tang vortex, as measured by $h \vert \nabla \cdot \textbf{B} \vert / \vert \textbf{B} \vert$, with and without divergence cleaning (orange dashed line and blue solid line, respectively). The shaded regions represent the inter-quartile range, that is, the 25$^{\rm th}$ and 75$^{\rm th}$ percentiles.}
\label{fig:ot-hdivbb}
\end{figure}

\subsection{Alternative divergence control approaches}

SPMHD calculations enjoy a minor level of divergence control through artificial resistivity. As discussed in section~\ref{sec:ar}, the explicitly added dissipation from artificial resistivity is equivalent to a (resolution-dependent) physical dissipation, $\eta_{\rm AR} \nabla^2 \textbf{B} = \eta_{\rm AR} \nabla (\nabla \cdot \textbf{B}) - \eta_{\rm AR} \nabla \times (\nabla \times \textbf{B})$. Though artificial resistivity is intended to treat discontinuities in the magnetic field, it may help dissipate magnetic monopoles. The caveat is that dissipation is also applied to physical portions of the field, so using artificial resistivity primarily as a divergence control measure has its drawback.

Periodically smoothing the magnetic field to remove small-scale noise has occasionally been suggested \citep{bot01, ds09, se15, tu-etal22}. However, the frequency of the smoothing procedure needs to remove small-scale noise without overly smoothing the field. Smoothing procedures are adhoc without any formal guarantees.

Formulations that yield a truly divergence-free magnetic field would be ideal. SPMHD appears to be far from achieving this currently (see discussion on the Euler potentials and vector potential in sections~\ref{sec:ep} and \ref{sec:vecp}). One avenue that is under-explored are projection methods \citep{bb80, toth00}. Considering an ``unclean'' magnetic field, $\textbf{B}^*$, it can be written in terms of its physical and unphysical components according to $\textbf{B}^* = \nabla \times {\bf A} + \nabla \phi$, where ${\bf A}$ is the vector potential and is the physical portion of the field (since the divergence of the curl is zero). From this, it can be stated that $\nabla \cdot \textbf{B}^* = \nabla^2 \phi$, and then by solving for $\phi$, the divergence-free magnetic field can be obtained from $\textbf{B} = \textbf{B}^*- \nabla \phi$. It is noteworthy to consider that projection methods are commonly used to obtain divergence-free velocity fields for the simulation of incompressible fluids in SPH \citep{cr99, ha07, lrs20}. 

\citet{pm05} tested projection methods in SPMHD, finding that they worked well for simple test problems. The most significant disadvantage is the computational cost in solving the elliptic set of equations. Using a tree may help improve efficiency, but would still be similar in cost to tree-based gravity solvers. Individual particle timesteps add further complication. It may be worth revisiting projection methods and testing further.

\section{Non-ideal MHD}
\label{sec:non-ideal}

Non-ideal MHD concerns plasmas that are only partially ionized. A population of neutral particles can introduce important deviations from the flux freezing condition of ideal MHD. In the context of star formation, these non-ideal effects may play a role in molecular cloud formation, disc fragmentation and braking, and the seeded magnetic field strength inside protostars \citep[see reviews by][]{pr19, hi19, mhg22}. 

Consider a partially ionized plasma consisting of electrons, ions, and neutrally charged particles. Ohmic resistivity refers to the impediment of free electron flow in conducting plasmas. Ambipolar diffusion is the process of neutral particles drifting through ions. The magnetic field is tied to the electrons and ions, thus there is a movement of mass that does not result in transport of the magnetic field. Both Ohmic resistivity and ambipolar diffusion are dissipative processes. The Hall effect, on the other hand, is not dissipative, but rather dispersive. For the Hall effect, ions or charged dust grains are collisionally coupled to the neutral species, but the electrons are not. The magnetic field is only transported by the electrons. See, for example, \citet{wn99, bw12b, bw12a, pr19, tsukamoto-etal22} for discussion on non-ideal effects in the context of star formation.

The three non-ideal effects of Ohmic dissipation, ambipolar diffusion, and the Hall effect have all been implemented into SPMHD \citep{tii13, wpa14} and have been used for a variety of simulations related to star formation \citep[e.g.,][]{tsukamoto-etal15b, wpb16, wbp19}. Generally speaking, the context in which non-ideal effects would be expected to play a substantial role is when the ionization rate is sufficiently low. For this reason, \citet{wpa14} implement non-ideal effects using a single species of SPMHD particles using the strong coupling approximation, though it is possible to model using two species of particles, e.g., ambipolar diffusion \citep{hw04}.

The continuum equations for the non-ideal MHD effects are given by
\begin{align}
\frac{{\rm d} \textbf{B}}{{\rm d}t} = & - \nabla \times \left[ \eta_{\rm O} (\nabla \times \textbf{B}) \right] \nonumber \\
& + \nabla \times \left\{ \eta_{\rm A} \left[ (\nabla \times \textbf{B}) \times \hat{\textbf{B}} \right] \times \hat{\textbf{B}} \right\} \nonumber \\
& - \nabla \times \left[ \eta_{\rm H} (\nabla \times \textbf{B}) \times \hat{\textbf{B}} \right] ,
\end{align}
where $\hat{\textbf{B}} = \textbf{B} / \vert \textbf{B} \vert$, that is, the unit vector in the direction of the magnetic field, and $\eta_{\rm O}$, $\eta_{\rm A}$, and $\eta_{\rm H}$ are the coefficients for Ohmic resistivity, ambipolar diffusion, and the Hall effect, respectively. 

Non-ideal MHD effects introduce additional timestep constraints for numerical stability. This is given by
\begin{equation}
\Delta t = \frac{C_{\eta} h^2}{\eta} ,
\end{equation}
which is calculated for each non-ideal effect. \citet{phantom} use $C_{\eta} = 1 / (2 \pi)$. This timestep constraint can become dominant for large $\eta$, and additionally note that it scales $\propto h^2$, whereas the Courant condition is only $\propto h$. Super-time-stepping may be of some benefit to alleviate pressure from the non-ideal MHD timestep constraints \citep{aag96, tii13, wpb16}, though this is only applicable for Ohmic dissipation and ambipolar diffusion.

\subsection{Ohmic dissipation}

\citet{tii13} and \citet{wpa14} implemented Ohmic resistivity in SPMHD using a `two-first derivatives' approach (also for ambipolar diffusion and the Hall effect). This approach makes including non-constant resistivity coefficients straightforward. The discretised equations are given by
\begin{equation}
\label{eq:curlb-diff}
\nabla \times \textbf{B}_a = \frac{1}{\Omega_a \rho_a} \sum_b m_b (\textbf{B}_a - \textbf{B}_b) \times \nabla W_{ab}(h_a) ,
\end{equation} 
\begin{align}
\frac{{\rm d}}{{\rm d}t} \left( \frac{\textbf{B}}{\rho} \right)_{a} = & - \sum_b \bigg[ \frac{\eta_{{\rm O},a} (\nabla \times \textbf{B})_a}{\Omega_a \rho_a^2} \times \nabla_a W_{ab}(h_a) \nonumber \\
& + \frac{\eta_{{\rm O},b} (\nabla \times \textbf{B})_b}{\Omega_b \rho_b^2} \times \nabla_a W_{ab}(h_b) \bigg] .
\end{align}
The curl of $\textbf{B}$ is calculated first using a difference derivative estimate, then the curl of the result with a symmetric derivative estimate. This conjugate pair of derivative operators appear in many contexts \citep[e.g.,][]{cr99, price10, tp12}. The Ohmic dissipation yields a positive-definite increase in thermal energy according to
\begin{equation}
\frac{{\rm d}u_a}{{\rm d}t} = \frac{\eta_{{\rm O},a} (\nabla \times \textbf{B})_a^2}{\rho_a}
\end{equation}

Note that artificial resistivity is equivalent to an Ohmic dissipation (see section~\ref{sec:ar}). In this case, it is calculated using a second derivative directly in a manner equivalent to \citet{brookshaw85} and \citep{cm99}, rather than with two first derivatives as above. Ohmic resistivity implemented with a direct second derivative does work well, provided the resistivity coefficient is spatially constant \citep{bonafede-etal11, bp19}.

\subsection{Ambipolar diffusion}

Ambipolar diffusion can be implemented using a two-first derivatives approach, with the inner derivative, $(\nabla \times \textbf{B})_a$, calculated using a difference derivative estimate (Equation~\ref{eq:curlb-diff}). The second derivative is estimated with a symmetric derivative, according to
\begin{align}
\label{eq:ad-symm}
\frac{{\rm d}}{{\rm d}t} \left( \frac{\textbf{B}}{\rho} \right)_{a} = &-  \sum_b \bigg[ \frac{{\bf D}_{{\rm A},a}}{\Omega_a \rho_a^2} \times \nabla_a W_{ab}(h_a) \nonumber \\
& \hspace{8mm}+ \frac{{\bf D}_{{\rm A}, b}}{\Omega_b \rho_b^2} \times \nabla_a W_{ab}(h_b) \bigg] ,
\end{align}
with
\begin{equation}
{\bf D}_{{\rm A},a} = - \eta_{{\rm A},a} \left[ (\nabla \times \textbf{B})_a \times \hat{\textbf{B}}_a \right] \times \hat{\textbf{B}}_a .
\end{equation}
A positive-definite increase in thermal energy is ensured \citep{wpa14}, with
\begin{equation}
\frac{{\rm d}u_a}{{\rm d}t} = \frac{\eta_{{\rm A},a}}{\rho_a} \left\{ (\nabla \times \textbf{B})_a^2 - \left[ (\nabla \times \textbf{B})_a \cdot \hat{\textbf{B}}_a \right]^2 \right\} .
\end{equation}

\subsection{Hall effect}

The Hall effect calculates $(\nabla \times \textbf{B})_a$ in the same manner as Ohmic resistivity and ambipolar diffusion, that is, according to the difference derivative estimate in Equation~(\ref{eq:curlb-diff}). The outer derivative is calculated using a symmetric derivative estimate according to Equation~(\ref{eq:ad-symm}), but with
\begin{equation}
{\bf D}_{{\rm H},a} = \eta_{{\rm H},a} (\nabla \times \textbf{B}_a) \times \hat{\textbf{B}}_a
\end{equation}
in place of ${\bf D}_{\rm A}$.

As the Hall effect is dispersive, and not dissipative, it only leads to the re-distribution of magnetic energy. There is no deposition into thermal energy.

The Hall effect introduces an additional wave type known as whistler modes \citep{ss02, pw08, bai14}. The left and right polarizations of Alfv\'{e}n waves become asymmetric in the presence of the Hall effect, that is, they have different phase velocities. The right-handed polarization is called the whistler wave. The introduction of an additional wave type can be generally understood in that the Hall effect is hyperbolic in nature rather than parabolic.

\section{Prospective Outlook}
\label{sec:future}

If there is any indication on the future potential of SPMHD as a tool to study astrophysical problems, one only needs to look at the large bodies of work that have been conducted over the past decade in a variety of astrophysical contexts. It may taken decades for SPMHD to reach its current level of maturity, but all major roadblocks have been cleared. SPMHD is finally a method that is generally applicable for the study of astrophysics. This is not to say that there is no room to improve the method further. Far from it. 

With respect to grid-based codes, SPMHD possesses a number of advantages, such as its conservation properties, adaptive resolution with respect to the density, and absence of dissipation due to advection. On the other hand, the numerical dissipation arising from artificial viscosity and resistivity is typically greater than that stemming from Reimann-based solvers in grid codes. SPMHD is also typically more computationally expensive than grid-based codes, owing to the number of neighbours under the kernel ($\sim 60$--200) and the need for a nearest neighbour search. The inherent particle `re-meshing' motions additionally incur a level of background noise in the velocity field, though this can be addressed by using more neighbours or better kernels, but at increased computational expense \citep[e.g.,][]{tricco19}. For magnetic fields, SPMHD can only approximately uphold the divergence-free constraint at present, whereas grid-based codes can utilize approaches such as constrained transport to exactly solve for a divergence-free magnetic field. Furthermore, the magnetic field can only be specified where there is mass present (i.e., particles).

One area that deserves more study is the MRI. \citet{deng-etal19} found that SPMHD adequately simulates the MRI in unstratified shearing boxes with net flux, but have decaying solutions for zero-net flux. They also found that SPMHD produces unphysical behaviour for stratified shearing boxes, though the calculations of \citet{wissing-etal22} were able to sustain MRI-induced turbulence in stratified shearing boxes for over 100 orbits, seemingly due to the geometric density average formulism (see Figure~\ref{fig:mri}). \citet{wissing-etal22} were also able to sustain turbulence in zero-nut flux unstratified shearing boxes, provided the magnetic Prandtl number was above a critical threshold. Part of the complication of studying the MRI is the breadth of initial configurations and parameters. One avenue worth exploring is the inclusion of physical dissipation to set the Reynolds and magnetic Reynolds numbers, along with the magnetic Prandtl number. Another avenue that seems ripe is global disc simulations. SPH has a rich history of simulating accretion discs owing to its conservation properties, and applying SPMHD in this direction would appear sensible.

\begin{figure}
\begin{center}
\includegraphics[width=85mm]{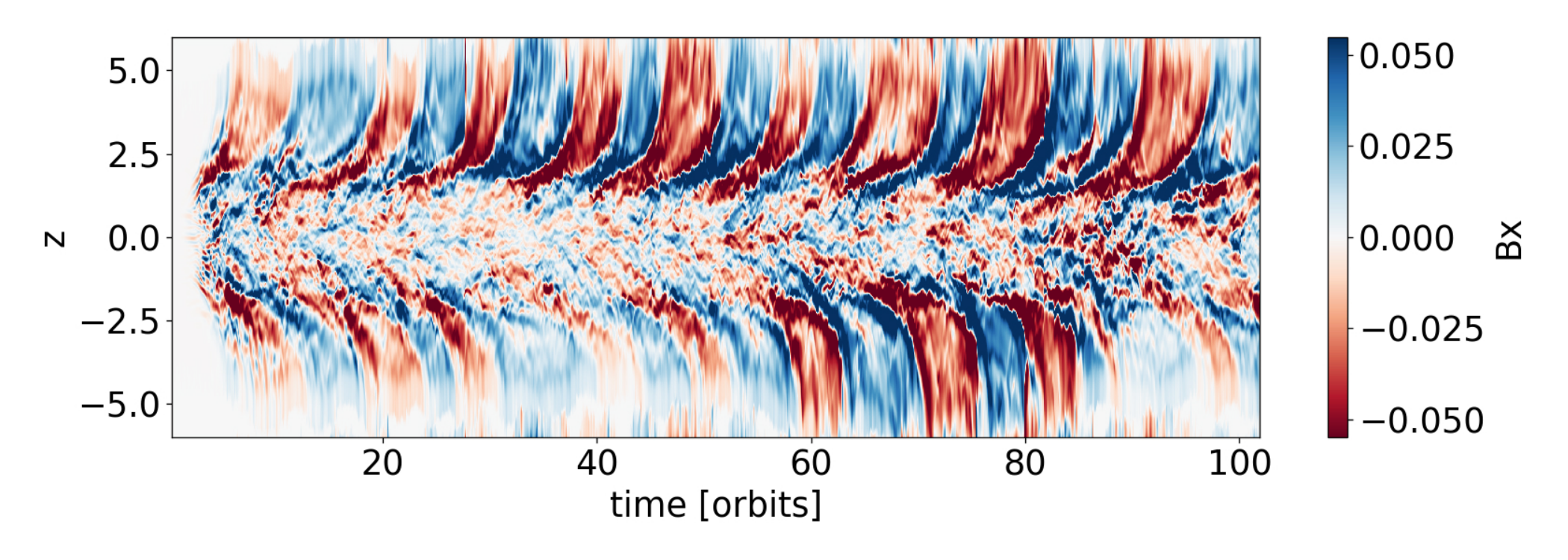}\\
\includegraphics[width=85mm]{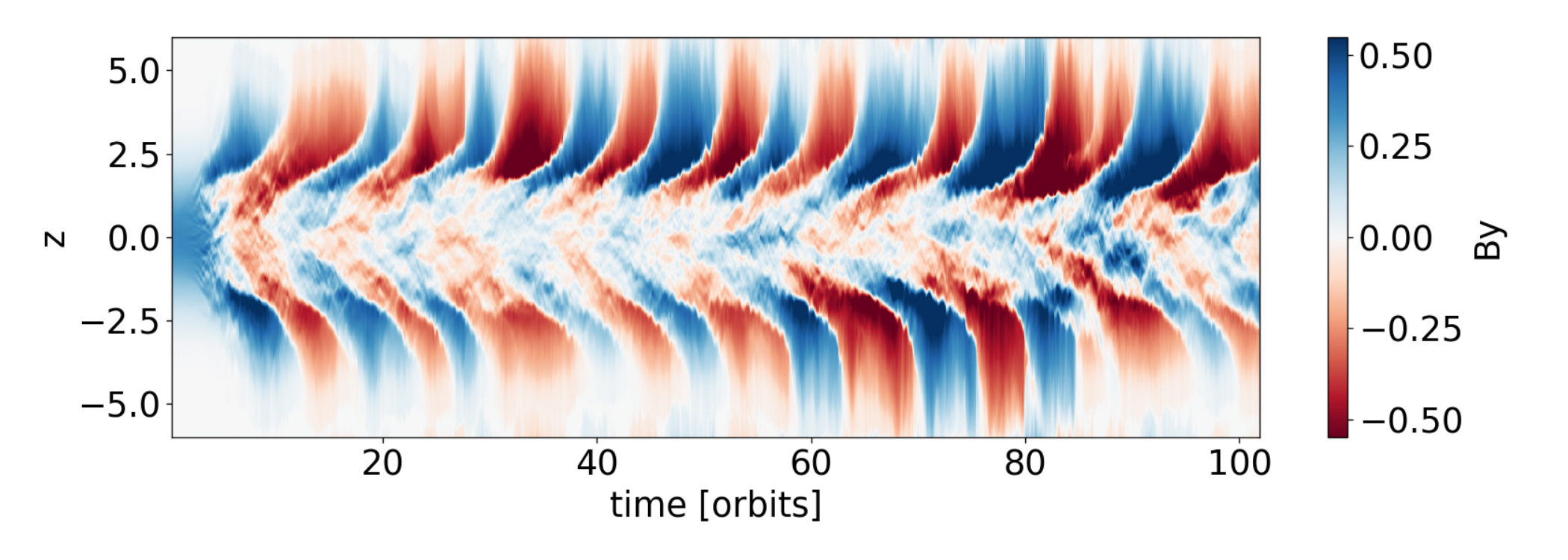}
\end{center}
\caption{Spacetime diagrams of magneto-rotational instability (MRI) calculations in a stratified net flux shearing box over 100 orbits. The top panel shows the horizontal averaged radial magnetic field, and the bottom panel the azimuthal magnetic field. The characteristic `butterfly' diagram is produced, whereby azimuthal fields are transported out of the plane of the disc and the direction of the magnetic field periodically reverses. Reproduced with permission from Figure~15 in \citet{wissing-etal22}.}
\label{fig:mri}
\end{figure}

There is also need to explore the properties of SPMHD on other fundamental MHD phenomena and to validate correctness of obtained solutions. Dynamos are one class of problem deserving more attention. To date, there has only been limited exploration of the small-scale dynamo \citep{tpf16} and galactic dynamos \citep{ws23}. And only the surface has yet been scratched on the properties of magnetised turbulence in SPMHD.

There should be continued effort to improve the fundamental accuracy or convergence of SPMHD. There are discussions and advancements within hydrodynamical schemes for Lagrangian, particle-based methods that may be relevant for SPMHD. \citet{gsce12} developed an SPH scheme that has higher-order gradient estimates through matrix inversion, and \citet{ii11} use Riemann solvers inside their Godunov-SPH scheme. 

Meshless finite mass (MFM) \citep{hopkins15, hr16} is another Godunov-type scheme that uses a least-squares matrix gradient operator and Riemann solvers with slope limiters. Like SPMHD, MFM uses mixed hyperbolic / parabolic divergence cleaning to uphold the divergence-free constraint on the magnetic field, but also employs a constrained-gradient method for further reduction of divergence errors in the magnetic field \citep{hopkins16}. At present, it is difficult the quantify how important is the choice of numerical scheme as there is a lack of overlap in astrophysical simulations between MFM and SPMHD, with use of MFM focused primarily on galaxy formation \citep[e.g., the FIRE simulations][]{fire1, fire2, fire3} and SPMHD on star formation calculations. Perhaps there may be lessons from these numerical methods, or even others, in regards to improvements for SPMHD.
 
Coupling magnetic fields with charged dust is a perfect opportunity to extend SPMHD in new directions. Non-ideal MHD effects and dust are tied together, as dust grains can adsorb electrons or ions. There are already robust solvers for gas-dust mixtures in SPH \citep{lp12a, lp12b, pl15}, and there is nothing in principle preventing creation of algorithms to couple magnetic fields and dust together in a unified solver. \citet{tmi21, tmi23} have made initial steps in this direction, expanding upon the one-fluid dust approach of \citet{lp14} to include magnetic fields. They make a number of simplifications, and better treatment on the differences between neutral and charged dust are still required (depending upon the physical regime).

Finally, it is worth noting that adapting existing SPH codes to include magnetic fields is straightforward. At its core, this involves evolving another variable (the magnetic field), adding magnetic dissipation terms, calculating accelerations from the magnetic field, and solving the divergence cleaning equations alongside the evolution of the magnetic field. While there are improvements that can still be made to the method, recent advancements and achievements with SPMHD suggest that it is capable of general theoretical studies of astrophysics, providing a complementary approach to grid-based methods.

\section*{Author Contributions}

The author confirms being the sole contributor of this work and has approved it for publication.

\section*{Funding}

TST was supported by a Discovery Grant from the Natural Sciences and Engineering Research Council (NSERC) of Canada. This research was enabled in part by support provided by ACENET and the Digital Research Alliance of Canada.

\section*{Acknowledgments}

TST thanks Daniel Price for many conversations over the years that has refined his understanding of, and appreciation for, SPH. TST thanks Robert Wissing for permission to reproduce the MRI figures. Figures and analysis in this review made use of \textsc{Sarracen}, a Python package for the analysis and visualization of SPH data \citep{sarracen}.

\bibliographystyle{Frontiers-Harvard} 
\bibliography{spmhd}

\end{document}